\documentclass[traditabstract]{aa}
\pdfoutput=1
\usepackage[varg]{txfonts}
\usepackage{supertabular}
\usepackage[bookmarks=false]{hyperref}
\usepackage{subcaption}
\usepackage{natbib}
\usepackage{booktabs}
\usepackage{longtable}
\usepackage{supertabular}
\usepackage{graphicx,verbatim}
\usepackage{txfonts}
\usepackage{natbib,twoopt}
\bibpunct{(}{)}{;}{a}{}{,} 

\def\lxlr{$P_{1.4\mathrm{GHz}}$ - $L_{X}$\,}

\begin{document}
%
   \title{The Extended GMRT Radio Halo Survey II: Further results and analysis of the full sample}

   \subtitle{}
\titlerunning{The Extended GMRT Radio Halo Survey II}
\authorrunning{Kale et al.}

   \author{R. Kale\inst{1,2,3}, T. Venturi\inst{1}, S. Giacintucci\inst{4,5}, D.
Dallacasa\inst{1,2}, R. Cassano\inst{1}, G.
Brunetti\inst{1}, V. Cuciti\inst{1,2}, G.
Macario\inst{6} and R. Athreya\inst{7}}
\institute{INAF-Istituto di Radioastronomia, via Gobetti 101, 40129 Bologna, Italy
\and Dipartimento di Fisica e Astronomia, Universita di Bologna, via
Ranzani 1, 40127 Bologna, Italy
\and National Centre for Radio Astrophysics, TIFR, Ganeshkhind, Pune - 411007, India \\
\email{ruta@ncra.tifr.res.in}
\and Department of Astronomy, University of Maryland, College Park, MD
20742, USA
\and Joint Space-Science Institute, University of Maryland, College Park,
MD, 20742-2421, USA
\and Laboratoire Lagrange, UMR7293, Universite de Nice Sophia-Antipolis, CNRS,
Observatoire de la Cote dAzur, 06300 Nice, France
\and Indian Institute of Science Education and Research (IISER), Pune, India }

   \date{A\&A accepted}

 
  \abstract{The intra-cluster medium contains cosmic rays and magnetic fields that are manifested through 
   the large scale synchrotron sources, termed as radio halos, relics and mini-halos. The 
    Extended Giant Metrewave Radio Telescope (GMRT) Radio Halo Survey (EGRHS) is an extension of the GMRT Radio 
    Halo Survey (GRHS) designed to search for radio halos 
    { using GMRT 610/235 MHz observations}. 
    The GRHS+EGRHS consists of 64 clusters in the redshift range 0.2 -- 0.4 that have an 
    X-ray luminosity larger than $5\times10^{44}$ 
    erg s$^{-1}$ in the 0.1 -- 2.4 keV band and with $\delta > -31^{\circ}$ in the REFLEX and eBCS X-ray cluster catalogues.
   In this second paper in the series, GMRT 610/235 MHz data on the last batch of 11 galaxy clusters 
   and the statistical analysis of the full sample are presented.
    A new mini-halo in RX\,J2129.6+0005 and candidate diffuse sources in 
    Z5247, A2552 and Z1953 are discovered. A unique feature of this survey are the upper limits on the detections 
    of 1 Mpc sized radio halos; 4 new are presented here making a total of 31 in the survey.
   Of the sample, 58 clusters that have adequately sensitive radio  
   information were used to obtain the most accurate occurrence fractions so far. 
   The occurrence of radio halos in our X-ray selected sample is $\sim22\%$, that of mini-halos is $\sim13\%$ 
   and that of relics is $\sim5\%$.
   The \lxlr diagrams for the radio halos and mini-halos with the detections and upper limits are presented.
   The morphological estimators namely, centroid shift ($w$), concentration parameter ($c$) and power ratios ($P_3/P_0$) 
   derived from the Chandra X-ray images are used as proxies for the dynamical states of the GRHS+EGRHS clusters. 
   The clusters with radio halos and mini-halos occupy distinct quadrants in the $c-w$, $c-P_3/P_0$ 
   and $w - P_3/P_0$ planes, corresponding to the more and less morphological disturbance, respectively. 
  The non-detections span both the quadrants. 
}

    \maketitle

\section{Introduction}\label{intro}
The intra-cluster medium (ICM) is the diffuse matter that pervades the space between the galaxies in  
galaxy clusters. It is dominated by hot ($\sim 10^7-10^8$ K) thermal plasma which emits thermal Bremsstrahlung 
detectable in soft X-ray bands and is also responsible for the thermal 
Sunyaev-Zel'dovich (SZ) effect. 
It has been found that the ICM also contains magnetic fields ($0.1-1 \mu$G) and relativistic electrons  
(Lorentz factors $>>1000$) distributed over the entire cluster volume. The most direct evidence for it 
are the cluster-wide diffuse synchrotron sources detected in radio bands \citep[see][for reviews]{fer12,bru14}. 
They occur in a variety of morphologies and sizes and are classified into three main types, namely, 
radio  halos, radio relics and mini-halos. 

Radio halos (RHs) are Mpc-sized sources found in massive, merging galaxy clusters. 
They typically trace the morphology of the X-ray surface brightness, show negligible polarization and have synchrotron 
spectral indices\footnote{The synchrotron spectral index, $\alpha$ is defined as $S_{\nu}\propto\nu^{-\alpha}$, 
where $S_\nu$ is the flux density at the frequency $\nu$.}, $\alpha \geq 1$ . The emitting relativistic electrons are 
believed to be 
reaccelerated via MHD turbulence \citep{bru01,pet01,pet08} although secondary electrons generated by 
hadronic collisions in the ICM  can also contribute \citep{den80,bla99,dol00,2011MNRAS.410..127B}. Turbulence can be 
injected in the ICM 
by dynamical activity, such as an infall of a sub-cluster in the cluster \citep[e.g.][]{sub06,ryu08}. Most of the 
massive and 
merging clusters are host 
to radio halos \citep[e.g.][]{cas13} however the turbulent reacceleration model 
also predicts a population of ultra-steep spectrum radio halos (USSRHs, $\alpha>1.5$ at classical frequencies 
$\sim1 $ GHz) from low energy cluster mergers.
The typical angular extents of the radio halos (a few arcminutes to 10s of arcminutes) and the predicted spectra make a 
strong case for their search at low frequencies.

Radio relics are the elongated, arc-like and polarized
radio sources occurring singly or in pairs at the peripheries of merging galaxy clusters; in a few cases associated with 
shocks detected in X-rays \citep[see][for a review]{bru14}. Their
 integrated radio spectra are typically steep ($\alpha>1$); however in a number of well studied cases, 
 a spectral index steepening trend from their outer to inner edges has been found  \citep{gia08,wee10,wee12,kal12}. 
 Detailed spectral analysis of the `sausage' relic has indicated possibilities of 
 particle reacceleration and/or efficient particle transport in the region behind the shock \citep{str14}.
 The understanding of the origin of seed relativistic electrons injected in the shock and the acceleration efficiency 
are still incomplete \citep[e.g.][]{mar05,kan11,kan12,pin13,vin14,vaz14,guo14,bru14}.

Mini-halos are diffuse radio sources found surrounding the dominant central galaxies in a number of cool core clusters. 
These have linear sizes of the order of a few hundreds of kpc.  
It has been shown that the ``sloshing'' of the gas in cluster cores can generate significant turbulence \citep{fuj04,zuh13}.
Furthermore, the spatial correlation between spiral cold fronts and edges of mini-halos seen in some cases 
\citep[e.g.][]{maz08,gia14} 
suggests that turbulence may be re-accelerating electrons 
to produce the mini-halos \citep{maz08, zuh13, 2014ApJ...795...73G}. In addition, the secondary electron models similar to 
those for radio halos, 
have also been suggested to explain the mini-halos that indeed trace the region where 
the number density of thermal targets in proton-proton collisions is higher \citep{pfr04,kes10,zan14}. 
Based on the role of turbulent motions in the ICM and the transport of cosmic rays a 
potential evolutionary connection between mini-halos and radio halos has also been 
suggested { \citep[e.g.][]{bru14, bon14}}.

A few dozen radio halos, relics and mini-halos have been found. 
The sensitivity and resolution to properly detect them require long observing time 
\citep[e.g.][]{has78,jaf79}. Therefore, targeted searches in the past 
have been limited to small samples.
Search for these sources in all sky surveys such as the NRAO VLA Sky Survey (NVSS) and the VLA Low-Frequency 
Sky Survey (VLSS) have led to discoveries of several 
new sources \citep{gio99,wee09}. 
 However, a tailored search for diffuse radio sources in a large sample of galaxy clusters was needed to make an objective 
 study of the statistics of their occurrence.
The steep spectra of radio halos and relics makes them suitable for search with a 
sensitive low frequency aperture synthesis instrument such as the Giant Metrewave Radio Telescope (GMRT).

The GMRT Radio Halo Survey (GRHS) was the first tailored search for radio halos 
at 610 MHz \citep[][hereafter V07 and V08, respectively]{ven07,ven08}. 
The results showed for the first time that clusters branched into two populations -- one 
with radio halos and another without -- that 
correlate with the dynamical state of the clusters \citep{cas10}.
This led to the understanding of the bimodality in the distribution of 
radio-loud and radio-quiet clusters in the radio power- X-ray luminosity plane (\lxlr) \citep{bru07}
in which an empirical correlation was known based on the known radio halos \citep{lia00}. 
The radio halos were found to be in $\sim 1/3$ of the 35 clusters for which radio information was available 
then.

The Planck satellite detections of galaxy clusters using the SZ effect and 
the GRHS initiated the exploration of the scaling between the radio halo power and 
the Compton Y-parameter \citep{bas12}. A weak or absent bimodality was reported
 in the $P_{1.4\mathrm{GHz}}- Y$ or equivalently in the $P_{1.4\mathrm{GHz}}- M$ 
plane. However with the recently released Planck SZ cluster catalog \citep{2014A&A...571A..29P} 
which is six times the size of the early cluster catalog,
the bimodality and its connection with mergers has been confirmed and quantified at least for 
the clusters with $M_{500}>5.5\times10^{14}\,M_{\odot}$ \citep{cas13}. 

In order to improve the statistics, the GRHS sample was extended to form the
Extended-GRHS (EGRHS). 
The observing strategy was also upgraded to dual-frequency (610 and 235 MHz).
The sample description and first results from the EGRHS were presented in \citet[][hereafter K13]{kal13}. 
The results included the detection of the mini-halo in RXCJ1532.9+3021 
at 610 MHz, radio halo upper limits on 10 and mini-halo upper 
limits on 5 clusters. The \lxlr \, diagram for cool-core clusters was also 
explored to investigate if there was bi-modality similar to that of radio halos.

In this paper, we present the radio data analysis of all the remaining clusters in EGRHS and the statistical 
results from the combined GRHS and EGRHS samples. In Sec.~\ref{sample} we describe the 
cluster sample. The radio observations and data reduction are described in Sec.~\ref{obs}.
The results are presented in Sec.~\ref{results} with the individual sources described in the subsections. 
The analysis of the GRHS+EGRHS sample is discussed in Sec.~\ref{disc2}. Summary of the work and conclusions are 
presented in Sec.~\ref{sum}.

A cosmology with $H_0 = 70$ km s$^{-1}$ Mpc$^{-1}$, $\Omega_m = 0.3$,  
and $\Omega_\Lambda = 0.7$ is adopted.

\section{The Sample}\label{sample}
The combined sample (GRHS+EGRHS) is extracted from the 
ROSAT-ESO flux-limited X-ray galaxy cluster catalog \citep[REFLEX,][]{boh04}
 and from the extended ROSAT Brightest Cluster Sample catalog
 \citep[eBCS,][]{ebe98,ebe00} based on the following criteria:
\vspace*{-0.2 cm}
\begin{enumerate}
 \item L$_X$ (0.1-2.4 keV) $> 5\times10^{44}$ erg s$^{-1}$;
 \item $0.2 < z < 0.4 $; and
 \item $\delta  > -31^{\circ}$.
\end{enumerate}
 At 610 MHz the GMRT has a resolution of $\sim 5''$ 
and is capable of imaging extended sources of up to $\sim 17'$. 
Surface brightness sensitivities $\sim 0.04 - 0.08$ mJy beam$^{-1}$ for extended sources
can be reached with $\sim 8$ hours of observing time. The redshift range of the sample 
is tailored so as to image linear scales $\sim 1$ Mpc with sufficient sensitivity and resolution to 
separate discrete sources. The sample also probes the redshift range where the 
bulk of the giant radio halos are expected to occur in the massive clusters according to theoretical models
\citep{cas06}. 

The GRHS was undertaken with a sample of 50 clusters (V07, V08) and the EGRHS with 17 clusters (K13). 
A combined sample of 64 clusters is formed after 
excluding three outliers \footnote{We remove the  clusters RXC\,J1512.2-2254, Abell 689 and RXJ2228.6+2037 
from the combined sample of 67 clusters reported until K13. 
The REFLEX catalogue source RXC\,J1512.2-2254 in the GRHS sample 
(V07, V08) contains no spectroscopically confirmed  
member-galaxies \citep{2009A&A...499..357G}. The X-ray luminosity of the cluster Abell 689 after subtraction of the central 
point source reported in \citep{gil12} 
 pushes it below our lower luminosity limit for selection. The redshift of RXJ2228.6+2037 is 
 above the survey cut-off of 0.4.}   (full table is presented 
in Tab.~\ref{tabonline}).

\begin{table*}[]
\caption[]{\label{t1} Properties of the clusters. The clusters in the first sector belong to the EGRHS sample (K13) and 
those in the second belong to the GRHS sample. Columns are 1. Cluster name; 2. Right ascension; 3. Declination; 4. Redshift;
5. X-ray
luminosity\tablefootmark{1} 
(0.1 -- 2.4 keV) in units $10^{44}$  erg s$^{-1}$ \citep{ebe98,ebe00,boh04}; 6
. Linear scale in $\mathrm{kpc}/''$; 
7. Morphological status (R= Relaxed and M= Merging)\tablefootmark{2}.}
\begin{center}
\begin{tabular}{llccccc}
\hline
\noalign{\smallskip}
Name&  RA$_{J2000}$ & DEC$_{J2000}$ & $z$ & L$_X$ (0.1 -- 2.4 keV)     & $\mathrm{kpc}/''$ & Morph. \\
\noalign{\smallskip}
    &  hh mm ss     & $^\circ  $\,\,$ ' $\,\,$ '' 	$  &   & $10^{44}$ erg s$^{-1}$ &        &status \\
\hline \noalign{\smallskip}
A68 &  00 36 59.4  &  +09 08 30     & 0.254 & 9.47  & 3.96 &  M\\
 \noalign{\smallskip}
Z1953 { (ZwCl0847.2+3617)}     &  08 50 10.1 &  +36 05 09 & 0.373 & 23.46\tablefootmark{3} &5.15 & M\\
 \noalign{\smallskip}
 Z3146 {(ZwCL1021.0+0426)}                     &  10 23 39.6 &  +04 11 10 & 0.290 & 17.26 & 4.35& R\\
  \noalign{\smallskip}
Z5247 {(RXC\,J1234.2+0947)}                   &  12 34 17.3 &  +09 46 12 & 0.229  & 6.32 & 3.66& M\\
 \noalign{\smallskip}
 A1722                    &  13 19 43.0 &  +70 06 17 & 0.327 & 10.78 &4.72 & R\\
  \noalign{\smallskip}
 RX\,J2129.6$+$0005	  &  21 29 37.9 &  +00 05 39 & 0.235  & 11.66 &3.73 &R\\
  \noalign{\smallskip}
  A2552	&  23 11 33.1 &  +03 38 07 & 0.305 &10.07 & 4.50&R\\
 \noalign{\smallskip}
\hline \noalign{\smallskip}
RXC\,J1212.3$-$1816        &  12 12 18.9 &  -18 16 43 & 0.269 & 6.20&4.12& M\tablefootmark{3}\\
 \noalign{\smallskip}
 A2485 & 22 48 32.9 & -16 06 23 & 0.247 & 5.10 &3.89 &R\\
  \noalign{\smallskip}
 RXC\,J1504.1-0248 &15 04 07.7 &-02 48 18& 0.2153& 28.08&3.50 &R \\
  \noalign{\smallskip}
 RXC\,J0510.7-0801& 05 10 44.7& –08 01 06 &0.2195& 8.55&3.55&M\\
\noalign{\smallskip}
\hline 
\end{tabular}
\tablefoot{\tablefoottext{1}{Corrected for cosmology.}
\tablefoottext{2}{\citet{cas10} and Sec.~\ref{clusdyn}}
\tablefoottext{3}{Possibly contaminated by point sources \citep{1999MNRAS.306..857C}.}
\tablefoottext{4}{Based on visual inspection of the XMM Newton image.}
}

\end{center}

\end{table*}

\section{Radio observations and data reduction}\label{obs}
The GRHS and EGRHS are large projects with radio observations 
 carried out in parts over several cycles of the GMRT. In this work we present the final batch of clusters 
 from the GRHS+EGRHS.
The following data are presented:
\begin{itemize}
 \item seven clusters from the EGRHS at 610 and 235 MHz,
\item four clusters from the GRHS for which there are new 610 and 235 MHz data.
\end{itemize}
The basic properties of these clusters are listed in Table~\ref{t1}.  
In addition to these, results on the GRHS clusters S780 and A3444 are 
briefly mentioned in this work but will be discussed in detail in a future paper 
(Giacintucci et al. in prep.).

Each of the { clusters} was observed in the 610-235 MHz (dual) band observing mode of the 
GMRT. {This mode allows simulataneous recording of data at two frequency bands but 
with a single polarization for each band, thereby reducing the effective sensitivity 
by $\sqrt{2}$.}
Bandwidths of 32 MHz at 610 MHz and 8 MHz at 235 MHz were used. The data were 
recorded with the GMRT Sofware Backend (GSB) in 256 frequency channels spread over 32 MHz.
Data analysis was described in detail in K13 and this work follows the same procedure. 
The data reduction was carried out in the Astronomical Image Processing System (AIPS). 
The standard data analysis s-eps-converted-to.pdf of flagging{(that includes removal of non-working antennas, bad baselines, 
radio frequency interference (RFI) affected channels and time ranges)}
and calibration were carried out on every dataset. The data on the target source were then separated 
and examined. These data were then averaged in frequency appropriately-- to reduce the computation 
time and at the same time to not be affected by the effect of bandwidth smearing. These 
visibilities were imaged to make an initial model for self-calibration. Several iterations of 
phase-only self-calibration were carried out. At a stage when most of the flux in the sources 
was cleaned, the clean components were subtracted from the uv-data. 
Low level RFI was excised from these data. The clean components were then added back and the data were used 
for further imaging and self-calibration. A final iteration of amplitude and phase self-calibration 
was carried out. Images with a range of high to low resolution using different weighting schemes 
and uv-range cutoffs were made to examine the cluster field for any diffuse emission. 
The rms noise in the images made with natural weights are reported in the Table~\ref{t2}.
High resolution images of discrete sources were made by excluding the inner uv-coverage 
(e.g. $< 1$ k$\lambda$). The low resolution images were made after subtracting the clean components of the 
high resolution images of discrete sources from the visibilities to examine diffuse emission on larger angular 
scales. The data quality was ensured to be high in most cases by carrying out night time observations as far 
as possible to minimise the RFI. Reobservations were carried out in cases of loss in observing duration 
due to power failure and scintillations or occurrence of more than 5 non-working antennas. 
However 4 datasets resulted in images whose quality was poorer than the average; 
these will be discussed under the respective sub-sections. 

The absolute flux density scale was set according to \citet{1977A&A....61...99B}.  
The residual amplitude errors ($\sigma_{amp}$) are in the range $5-8\%$ at 610 MHz and $8-10\%$ at 
235 MHz \citep[e.g.][]{2004ApJ...612..974C}. {The reported error ($\Delta S$) on the flux density ($S$) 
of a source is a combination of the rms noise in the radio image and the residual amplitude 
error expressed as, $\Delta S = \sqrt{(rms^2\times N_{beam}) + (\sigma_{amp}\times S)^2}$ , where 
$N_{beam}$ is the number of synthesized beams (the image resolution elements) 
within the extent of the source.}

\begin{center}
\begin{table}[]
 \caption{\label{t2} Summary of the GMRT images.}
\begin{tabular}{llll}
\hline   \noalign{\smallskip}
Cluster&Freq.&Beam & rms$^{\dag}$ \\
  \noalign{\smallskip}
 &MHz&$''\times''$, p. a. ($^{\circ}$) & mJy b$^{-1}$ \\
\hline   \noalign{\smallskip}
A68 &610& $9.3\times6.5$, $48.0$ & 0.05 \\
		  &235 & $29.1\times12.7$, $52.2$ &1.5\\
Z1953 		 &610&$7.6\times6.2$, $-63.6$ & 0.08 \\
              &235 & $17.8\times15.9 $, $-14.7$&0.5\\
Z3146 	      & 610 &$8.9\times7.1$, $11.3$ &0.09 \\
               &235 & $26.4\times22.9$, $44.9$ &1.1\\
A1722 	& 610&$8.6\times6.8$, $71.1$&0.05\\
		& 235&$17.4\times15.0$, $-65.9$&0.9\\
RXC\,J1212.3-1816 &610 &$11.4\times5.7$, 47.6 &0.07 \\
RX\,J2129.6+0005 &610 &$7.9\times6.3$, 29.3  &0.08  \\
		&235 &$29.7\times25.3$, 46.3 & 1.2\\
RXC\,J1504.1-0248 &610&$14.6\times7.7$, -13.7 &0.07\\
		&235 & $15.0\times11.8 $, 25.3 & 0.6\\
Z5247	&610	&$6.2\times5.2$, $56$	&0.03\\
       &235	&$13.0\times12.3$, $71.4$&0.6\\
A2552	&610	&$9.2\times7.2$, $50.4$	&0.07\\
		  	&235	&$16.2\times13.4$, $9.6$&0.7\\		  	
A2485	&610	&$8.1\times7.0$, $9.2$	&0.06\\
	&235	&$14.9\times11.5$, $50.3$& 1.0\\
RXC\,J0510.7-0801 & 610 &$5.4\times4.8$ &0.2$^{\ddag}$\\
		  &235 &$15.7\times13.1$ &1.2$^{\ddag}$\\
\hline
\end{tabular}
$^{\dag}$ Rms noise in images with natural weights.\\
$^{\ddag}$ {Rms noise in images with uniform weights.}
\end{table} 

\end{center}

\section{Radio images}\label{results}
The final images have typical rms noise ($1\sigma$) 
$\sim 30-80\, \mu$Jy beam$^{-1}$ at 610 MHz and 0.5 -- 1.2 mJy beam$^{-1}$ at 235 MHz (Tab.~\ref{t2}). 
An album of all the cluster fields of the size of the virial radius is presented in App.~\ref{appfig} and the 
images of the central regions of the clusters are discussed in the following sub-sections.
{ Flux density values for the diffuse and discrete sources in those clusters and 
their spectral indices are reported in Tab.~\ref{t3}. We note that the spectral 
indices of extended sources derived using flux densities at two frequencies may be affected 
by the differences in the uv-coverages at the two frequencies.}

\subsection{New detections}{\label{newdetect}}

\subsubsection{Z5247: A relic and a candidate halo}

\begin{figure*}
\centering
\includegraphics[height=8cm]{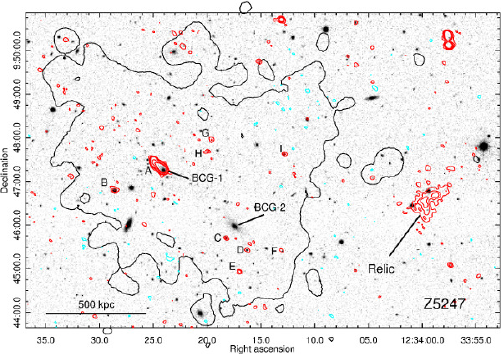}\\
\includegraphics[height=7cm]{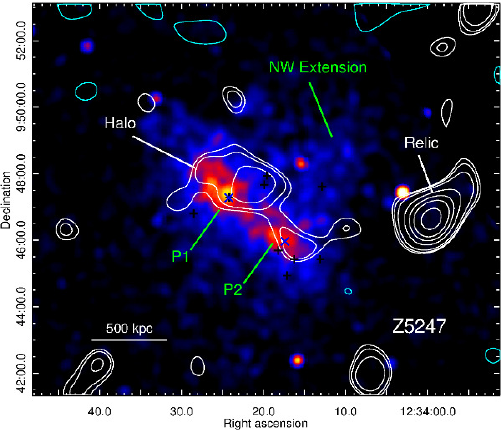}
\includegraphics[height=7cm]{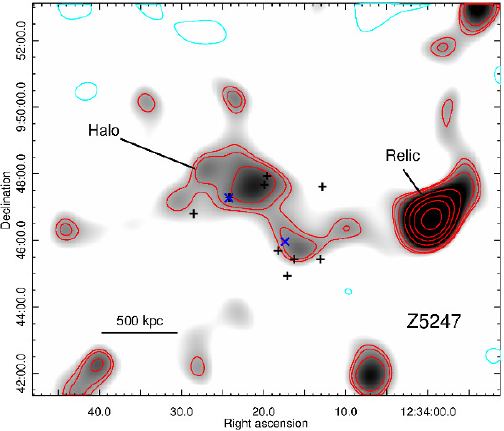}
\caption{{\bf Z5247}: {\it Top-} GMRT 610 MHz high resolution image in 
contours (red $+$ve and cyan $-ve$) overlaid on SDSS r-band image in grey-scale. 
The contours are at $-0.09, 0.09, 0.18, 0.36, 0.72, 1.44, 2.88, 5.76 $ mJy beam$^{-1}$. 
The beam at 610 MHz is  $6.2''\times5.2''$, p. a. $56^{\circ}$. The discrete 
radio sources in the cluster region (A to I), the relic and the two BCGs (1 and 2) 
are labelled. The extent of the X-ray emission from the cluster ICM 
is shown by the black contour. 
{\it Bottom left-} The Chandra X-ray image ({Obs ID 11727, 0.5 - 2 keV and resolution $\sim4''$}) of Z5247 
 is shown in colour overlaid with 
the contours of the $60''\times60''$ resolution 610 MHz image (black $+$ve and cyan $-$ve) 
made after the subtraction of the discrete sources. 
The contours are at $-0.9, 0.9, 1, 1.2, 1.8, 2.4, 3, 4$ mJy beam$^{-1}$. 
The features in the X-rays are labelled in green (P1, P2 and NW Extension) and 
the positions of the BCGs (`$\times$') and of 
the discrete radio sources (`$+$') are marked. 
{\it Bottom right-} 
 The $60''\times60''$ resolution 610 MHz image is shown in grey-scale with the same contours as in the left panel.
}
\label{z5247}
\end{figure*}

Z5247 (\object{RXC\,J1234.2+0947}) is an unrelaxed cluster with two BCGs (BCG-1, 2), each 
 corresponding to a peak (P1, P2) in the X-ray surface brightness distribution (Fig.~\ref{z5247}).
The BCG-1 is identified with a tailed radio galaxy (A in Fig.~\ref{z5247}, top panel).
An elongated diffuse source (relic, Fig.~\ref{z5247}) 
is detected at the low brightness edge of the X-ray emission (NW-Extension) about 1.3 Mpc to the west 
of the X-ray peak P1 (Fig.~\ref{z5247}).  The high resolution 
image at 610 MHz confirms the diffuse nature of this source. There are a few optical 
galaxies in the region however none of them is an obvious optical counterpart 
for the radio emission. 
If at the redshift of Z5247, the relic
 has a length of $\sim 690$ kpc and a maximum width of $\sim 320$ kpc detected in the low 
 resolution 610 MHz image. 

In addition, we detect diffuse radio emission 
in the central region of the cluster at 610 MHz (Fig.~\ref{z5247}, bottom panels). 
It is present over an extent of $\sim 930$ kpc 
along the northeast-southwest axis of the X-ray emission. We classify it as a candidate radio halo. 
It has a flux density of $\sim 7.9$ mJy at 610 MHz (Tab.~\ref{t3}).
We imaged the archival VLA D-array data at 1400 MHz (AG639) on this cluster and obtained 
an image with an rms noise of $\sim85 \mu$Jy beam$^{-1}$ (beam $65''\times45''$). 
The radio relic is detected and has a flux density of 3.1 mJy. The flux density of the central diffuse emission 
is difficult to estimate due to the presence of the embedded tailed galaxy. 
We estimated the flux density of A using high resolution FIRST survey image (1.4 GHz, B-array, resolution $\sim 5''$),
and obtained a residual flux density of $\sim 2$ mJy, which implies a spectral index of $\sim 1.7$ between 
610 and 1400 MHz. This value should be considered a lower limit as some extended emission associated with A might be 
missed in the FIRST image. The 235 MHz surface brightness sensitivity in lower resolution images was not 
sufficient to confirm the radio halo (rms noise of $\sim 2.3$ mJy beam$^{-1}$, beam $\sim60''$).

\subsubsection{RXJ2129.6+0005: A mini-halo}
\begin{figure*}
\centering
\includegraphics[height=6cm]{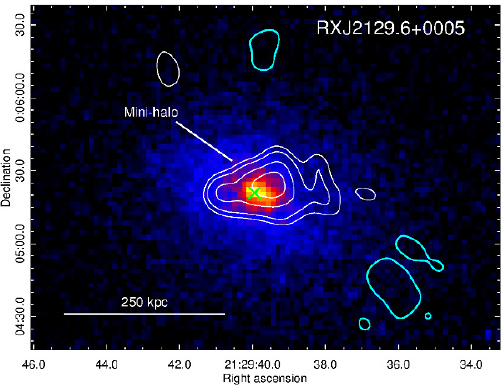}
\includegraphics[height=6cm]{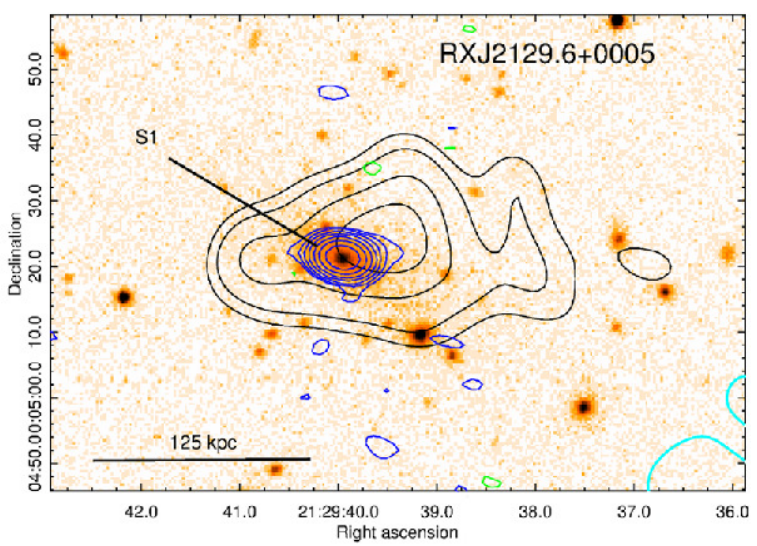}
\caption{{\bf RX\,J2129.6+0005}: {\it Left-} GMRT 610 MHz low resolution image in 
contours (white +ve and cyan -ve) is overlaid on the 
Chandra X-ray image ({Obs ID 09370, 0.5 - 2 keV and resolution $\sim2''$}) in colour. 
The contours are at $-0.11, 0.33, 0.44, 0.55, 0.66, 0.88$ mJy beam$^{-1}$ and the beam 
is $11.6''\times10.6''$, p. a. $28^{\circ}$. The compact source at the center marked by`$\times$' 
has been subtracted.
{\it Right-} The 610 MHz high resolution image in coutours (blue +ve and green -ve) is overlaid on 
the SDSS r-band image in colour. 
  The contours are at $-0.3, 0.3, 0.6, 0.12, 0.24, 0.48$ mJy beam$^{-1}$ and the beam is 
$9.3''\times6.5''$, p. a. $48^{\circ}$. The central compact source S1 is identified with the BCG.
 The contours from the left panel are also shown (black and cyan).
}
\label{r2129}
\end{figure*}

\begin{figure}
\centering
\includegraphics[height=6cm]{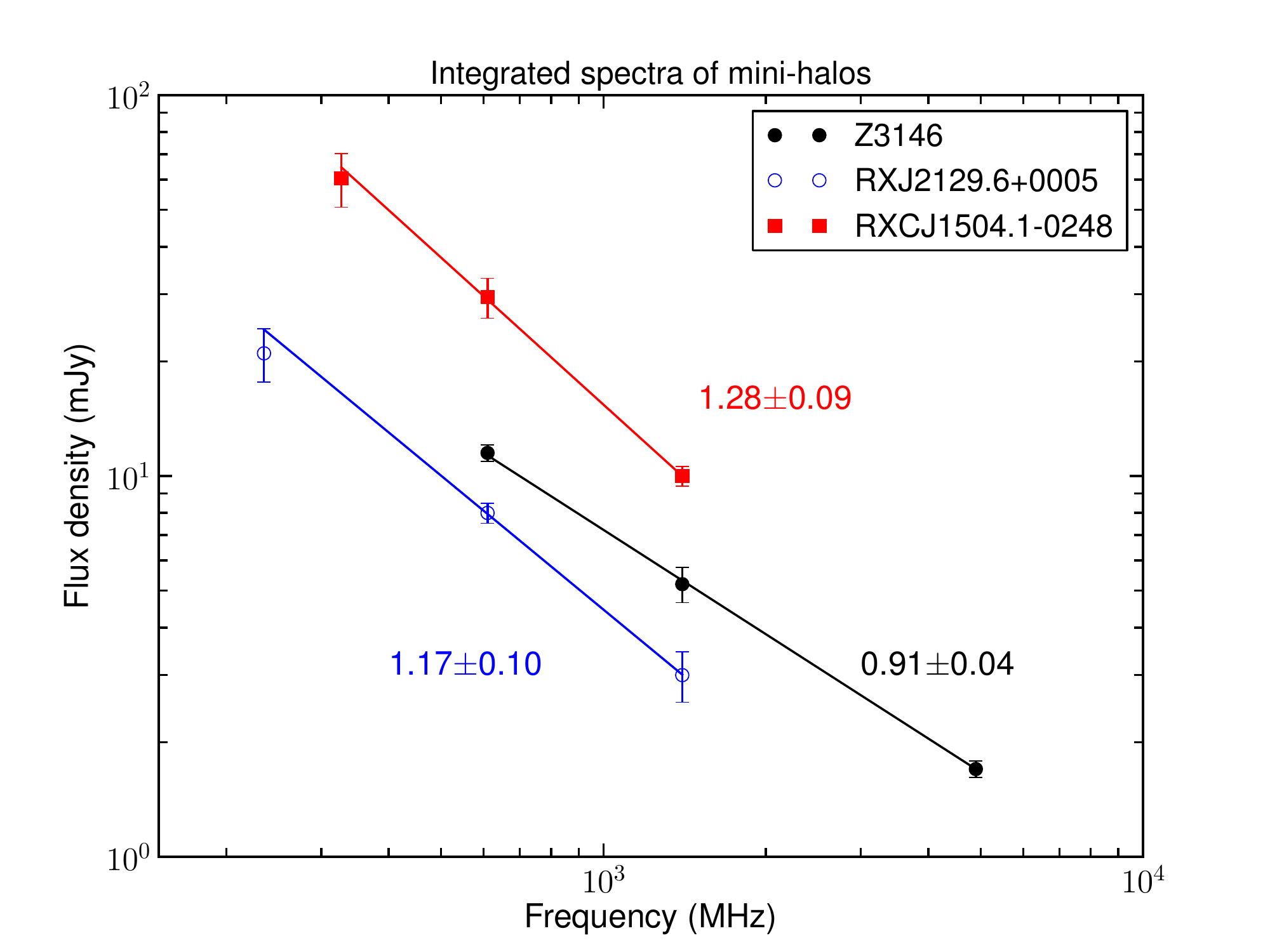}
\caption{The integrated spectra of the mini-halos in RXJ2129.6+0005, Z3146 and RXCJ1504.1-0248 are plotted. 
The best fit lines and the inferred spectral indices of the mini-halos are reported. 
The spectrum of RXCJ1504.1-0248 is scaled by a factor of 0.5 for visualization in this plot.}
\label{mhspec}
\end{figure}

\object{RXJ2129.6+0005} is a relaxed cluster with a 
BCG at the center hosting a strong radio source (Fig.~\ref{r2129}). The radio source 
is unresolved at the highest resolution in our GMRT 610 MHz observation and has a flux density 
of 48.9 mJy. 
This central source is surrounded by a low brightness extended emission that can be classified as a mini-halo
(Fig.~\ref{r2129}). It has a flux density of $\sim$8.0 mJy at 610 MHz and an extent of 225 kpc ($\sim 1'$) 
in the east-west and 110 kpc ($\sim 30''$) in the north-south direction. 
The flux density of the mini-halo at 235 MHz is 21 mJy. 
The residual flux density in the NVSS image after the subtraction of the flux density of discrete source from the 
FIRST image implies a flux density of 2.4 mJy for the mini-halo at 1400 MHz. 
The mini-halo has a spectral index of $\sim1.2$ over the frequency range of 
235 - 1400 MHz (Fig.~\ref{mhspec}).

\subsubsection{Z1953: A diffuse source}

\begin{figure*}
\centering
\includegraphics[height=8cm]{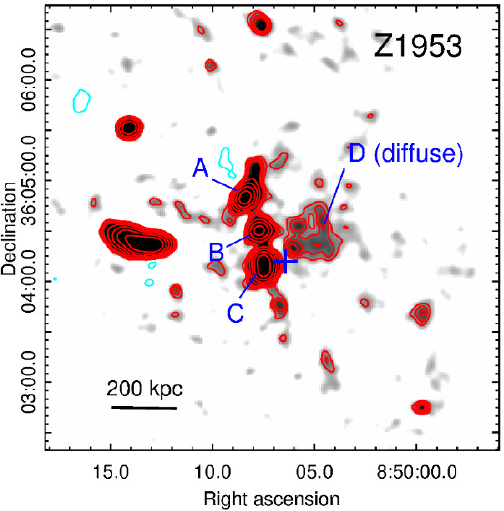}
\includegraphics[height=8cm]{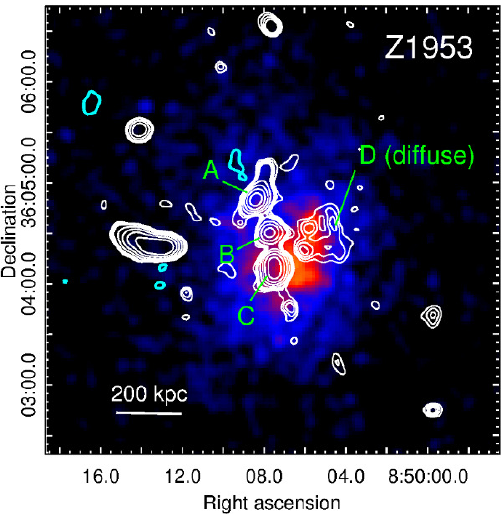}
\caption{{\bf Z1953:}{ \it Left-} GMRT 610 MHz image is shown in grey-scale 
and in contours. The contours are at $-0.15, 0.15, 0.2, 0.25, 0.3, 0.6, 1.2, 2.4, 4.8$ mJy beam$^{-1}$ and the 
beam is $9.3''\times6.5''$, p. a. $48^{\circ}$. The `$+$' marks the peak of the 
X-ray surface brightness. The discrete sources and the diffuse source are labelled. 
{\it Right-} 
The Chandra X-ray image ({Obs ID 01659, 0.5 - 2 keV and resolution $\sim4''$}) in colour overlaid with the 610 MHz contours from the left panel 
 (white +ve and cyan -ve).
}
\label{z1953}
\end{figure*}

Z1953 (\object{ZwCl\,0847.2+3617}) is a hot cluster classified as a merger based on optical 
and X-ray properties \citep{man12}. Three discrete radio sources (A, B and C) are located 
in the central region of the cluster, aligned in the north-south direction (Fig.~\ref{z1953}). 
To the west of these sources is a diffuse blob-like source, D. 
The source D has a roughly circular morphology with a diameter of $\sim 210$ kpc 
if assumed to be at the redshift of the Z1953. There is no obvious optical counterpart to D.
It has a flux density of 5.2 mJy at 610 MHz. It is too small in size to be classified 
as a candidate radio halo and its position is offset from the central BCG, unlike typical mini-halos. 
The dynamical activity in the cluster and its position make it an unusual mini-halo candidate. 
{ Thus we consider this to be unclassified diffuse emission.}

\subsubsection{A2552: A radio halo ?}
\begin{figure*}
\centering
\includegraphics[height=5.1cm]{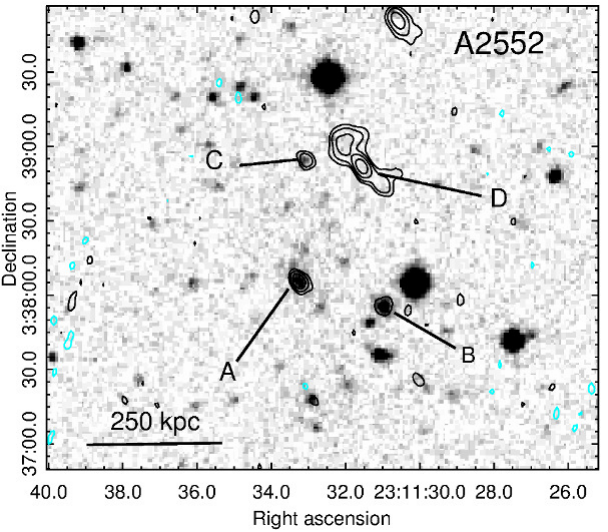}
\includegraphics[height=5.1cm]{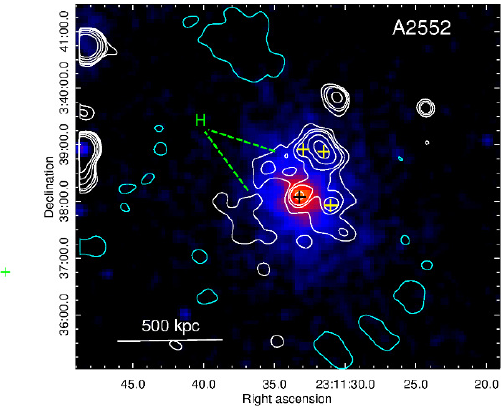}
\includegraphics[height=5.1cm]{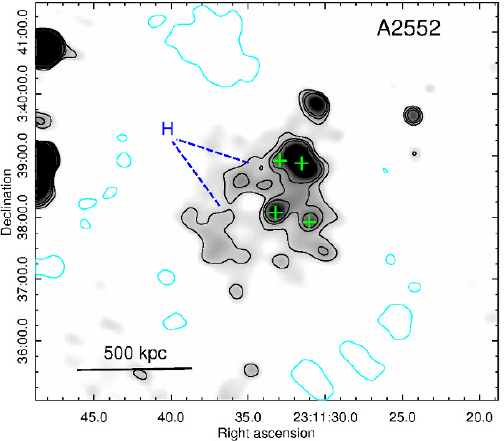}
\caption{{\bf A2552}: {\it Left-} GMRT 610 MHz high resolution image in contours overlaid on 
the DSS POSS II R band image in grey-scale. The contours are at 
$-0.15, 0.15, 0.30, 0.60, 1.2 $ mJy beam$^{-1}$ and the beam is 
$5.5''\times4.3''$, p. a. $35.9^{\circ}$.  
{\it Middle-} The 610 MHz low resolution (LR, $15''\times15''$) 
image in contours (white +ve and cyan -ve) 
overlaid on the Chandra X-ray image ({Obs ID 11730, 0.5 - 2 keV and resolution $\sim4''$}) in colour. 
The contours are at $0.4, 0.6, 0.8, 1.6, 3.2, 6.4$ mJy beam$^{-1}$.
The discrete sources are marked by `$+$' and H points out the patches of diffuse emission. 
{\it Right-} The 610 MHz LR image in grey-scale and the contours same as in the middle 
panel (black +ve and cyan -ve) are overlaid. 
}
\label{rxcj2311}
\end{figure*}

The cluster A2552 (RXC\,J2311.5+0338, \object{MACS\,J2311.5+0338}) is at a redshift of 0.305 \citep{ebe10}.
Four discrete sources are detected in the cluster central region at 610 MHz (Fig.~\ref{rxcj2311}, left):
A is associated with the {Brightest Cluster Galaxy (BCG) in the cluster and} 
B and C have optical counterparts. The morphology of source D 
is remeniscent of a lobed radio galaxy , however the the lack of optical identification makes its 
classification uncertain. If assumed at the redshift of A2552, D has a largest linear extent of 
180 kpc and a radio power of $2.6\times10^{24}$ W Hz$^{-1}$ at 610 MHz.

Patches of diffuse radio emission are detected in the cluster, 
around the central source A (labelled H, Fig.~\ref{rxcj2311}, center and right). 
The total flux density in the diffuse patches is $\sim 10$ mJy. However in the image obtained after the 
subtraction of clean components  of discrete sources, a flux density of $\sim 4.3$ mJy is found. 
This is nearly at the level of the typical upper limits obtained in 
our work at 610 MHz. 
Although classified as relaxed cluster based on the morphological parameter 
estimators, this cluster has been clssified as a `non cool-core' cluster based on X-ray and optical 
observations \citep{ebe10,lan13}. We classify the diffuse emission in A2552 as a candidate radio halo. 
The sensitivity at 235 MHz is not sufficient to 
detect the emission.
Deeper observations are necessary to confirm it.

\subsubsection{A3444 and S780}

\begin{figure*}
\centering
\includegraphics[height=8cm]{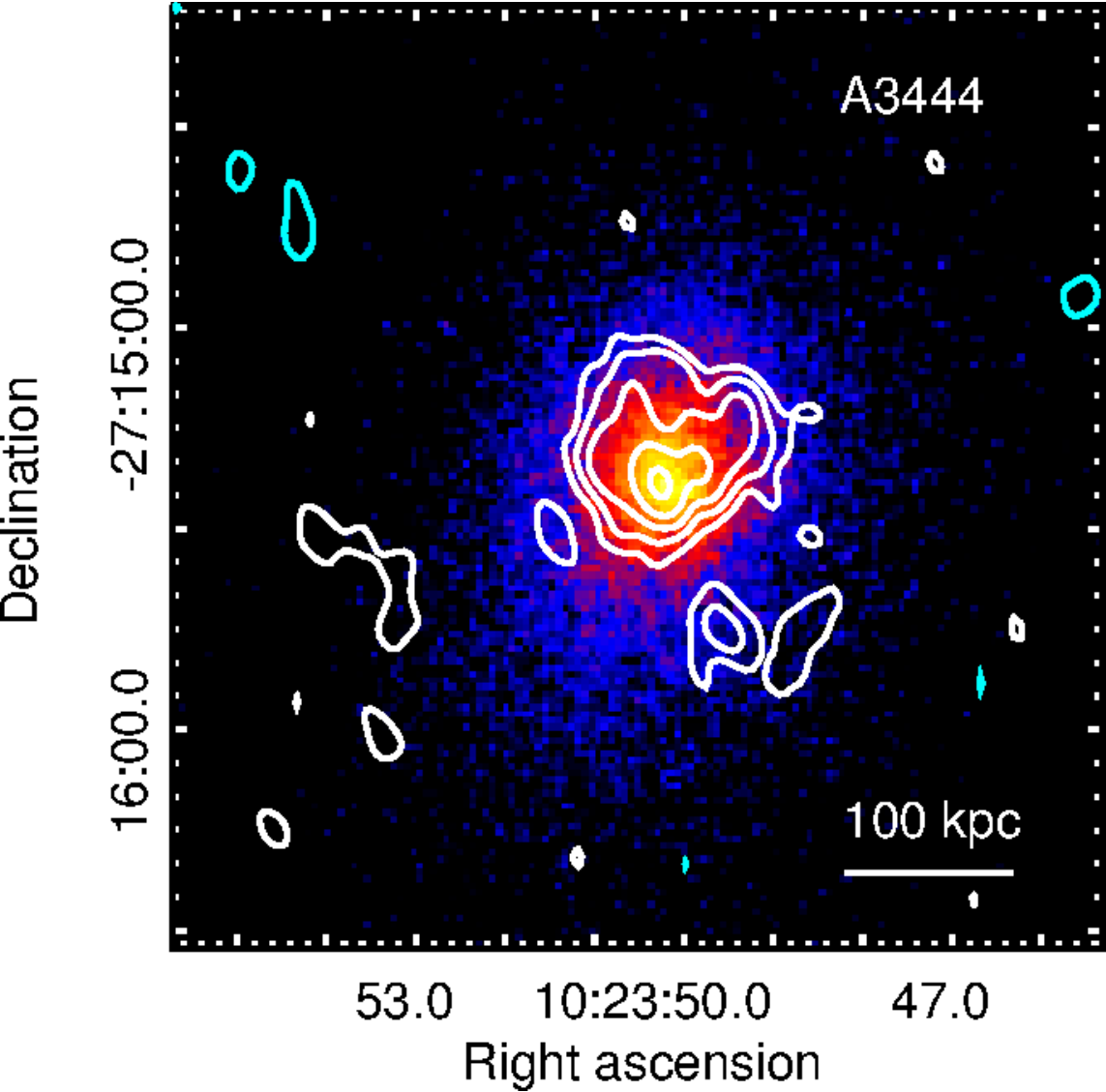}
\includegraphics[height=8cm]{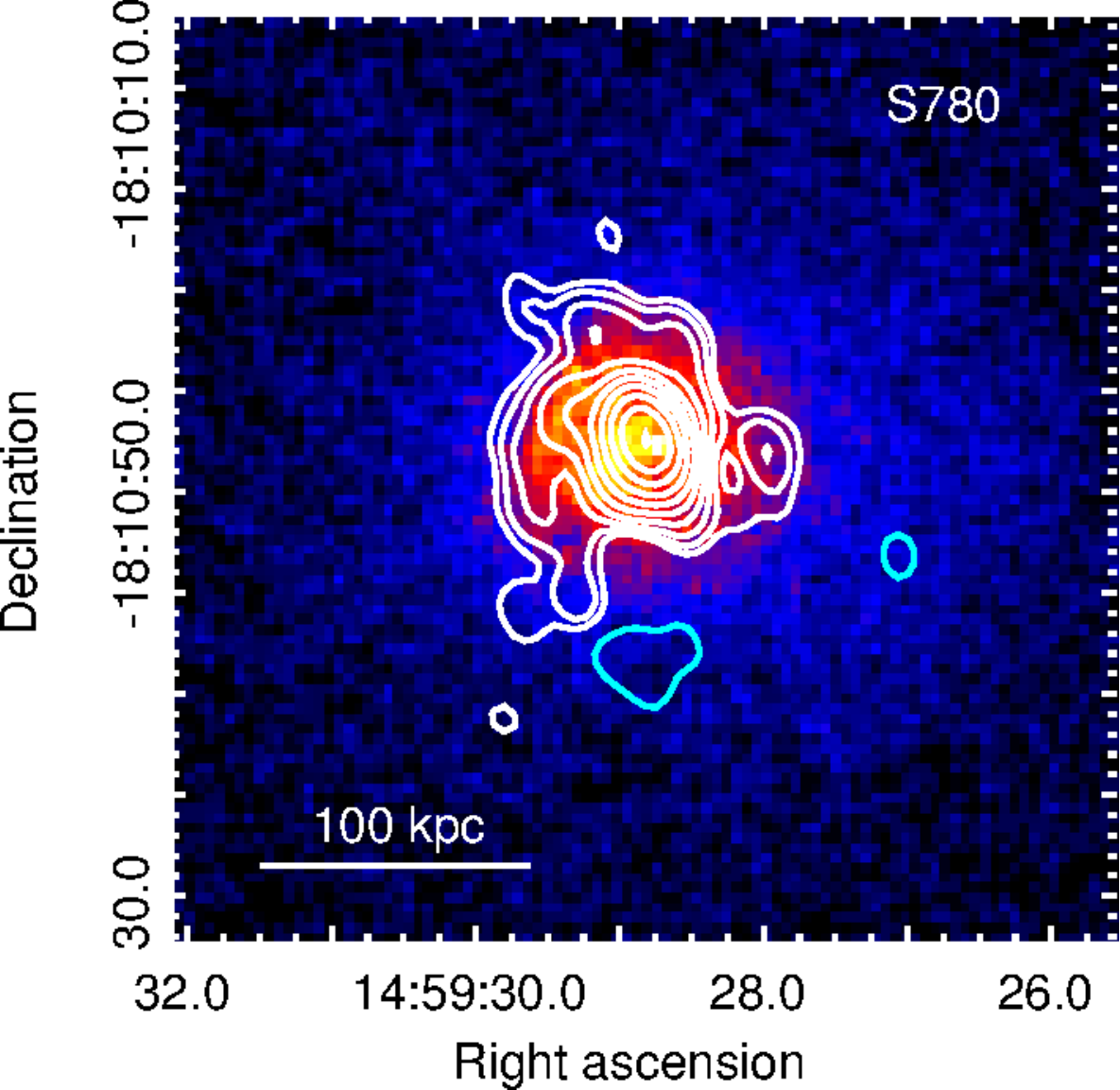}
  \caption{{\bf A3444 and S780} {\it Left-} GMRT 610 MHz
 image of A3444 from V07 in contours (white +ve and cyan -ve) 
overlaid on the Chandra X-ray image (Obs ID 9400, {0.5 - 4 keV and resolution $\sim 1''$}) in colour. 
The contours are at $0.2\times(\pm1, 2, 4,...)$ mJy beam$^{-1}$ and the beam 
is $7.6''\times4.9''$, p. a. $19^{\circ}$.
{\it Right-} GMRT 610 MHz image of S780 from the data presented in 
V07 shown in contours  (white +ve and cyan -ve) 
overlaid on the Chandra X-ray image (Obs ID 9428, {0.5 - 4 keV and resolution of $\sim1''$}) in colour. 
The contours are at $0.2\times(\pm1, 2, 4,...)$ mJy beam$^{-1}$ and 
the beam is $6''\times4''$.}
  \label{a3444}
\end{figure*}

A\,3444 and S\,780 were observed in the first part of the survey,
and presented in V07. A\,3444 was classified as a candidate
mini-halo, while S\,780 was considered as a non-detection,
despite the hints of diffuse emission surrounding the central radio
galaxy. More recently, Giacintucci et al. (in prep.)
started a statistical investigation of the properties of mini-halos
and their host clusters, and revised all the GMRT non-detections
and candidates, with analysis of archival VLA data at 1.4 GHz and
of $Chandra$ archival observations, not available at the time V07
was published. The good spatial coincidence of the X--ray 
features (cold front in S780 and an edge in A3444) with the radio 
morphology derived from the archival 1.4
GHz data and further imaging of the 610 MHz observations, led to
the classification of both clusters as mini-halos (Fig.~\ref{a3444}). 
Further new images and the mini-halo properties will be 
discussed in a future paper (Giacintucci et al.,
in prep.). 

\begin{table*}[]
\caption{\label{t3} Diffuse and discrete sources in selected cluster fields. The columns are: 1. Cluster name; 
2. Source label; 3. Right Ascension; 4. Declination; 5. Frequency; 6. Flux density; 7. Spectral index\tablefootmark{a}; 
8. Largest Linear Size; 9. Comments. }
\centering
\begin{tabular}{lcllccccr}
\hline\noalign{\smallskip}
Cluster		&Source	&RA$_{J2000}$ & DEC$_{J2000}$&$\nu$	&$S_\nu$&$\alpha$&LLS &Comments\\
Name		&	&  hh mm ss     & $^\circ  $\,\,$ ' $\,\,$ ''$ 	&MHz	&mJy	&	&kpc&\\
\noalign{\smallskip}
\hline
\hline\noalign{\smallskip}
{ Z5247} &	A	&12 34 24.2&$+09$ 47 18.3&\phantom{0}235	&$80\pm4$	&0.9	&&Radio galaxy\\
      &		&&&\phantom{0}610	&$35\pm2$	&$0.9$	&135&\\      
      &		&&&1400	&$16.4\pm0.8$	&	&&FIRST\\\cmidrule{2-9}
      \noalign{\smallskip}
      &Halo (H)	&12 34 21.1&$+09$ 47 21.2&\phantom{0}610 &$7.9\pm1$	&$1.7$	&930&Candidate\\
      &		&&&1400	&$2$	&	& &VLA-D -- FIRST\\
      &Relic (R)&12 33 59.6&$+09$ 46 33.9 	 &\phantom{0}235	&$36.8\pm3.0$	& 1.4&&\\
      &		&&&\phantom{0}610 &$9.3\pm1.0$	&		$1.3$&690&\\
      &		&&&1400	&$3.1\pm0.2$	&	&&VLA-D\\
\noalign{\smallskip}
\hline\noalign{\smallskip}
{ RX\,J2129.6+0005}	&S1	&21 29 39.9&$+00$ 05 21.3&235		&$87\pm5$	&$0.6$	&&BCG\\
\noalign{\smallskip}
				&&&	&610	&$48.9\pm2.5$	&$0.9$	&&\\
				\noalign{\smallskip}
				&&&	&1400		&$24.0\pm1.2$	&0.8	&&FIRST\\	
				\noalign{\smallskip}
				&&&	&4890		&8.9	&	&&VLA-C\tablefootmark{c}\\\cmidrule{2-9}
				\noalign{\smallskip}
			&Mini-halo&21 29 39.9&$+00$ 05 21.3	&\phantom{0}235	&$21.0\pm1.6$\tablefootmark{d}&$1.2$ &&$\alpha$ from Fig.~\ref{mhspec}\\
				\noalign{\smallskip}
			&&&		&\phantom{0}610	&$8.0\pm0.7$	&		&225&\\
			&&&	&1400		&$2.4$\tablefootmark{e}	&&& NVSS -- FIRST\\
				\noalign{\smallskip}
\hline\noalign{\smallskip}
{ Z1953}		&A	&08 50 08.4&$+36$ 04 50.1&\phantom{0}610	&$13.0\pm0.7$	&1.3	&&Tailed galaxy\\
			& &&	&1400		&$4.5\pm0.2$	&	&&FIRST\\
		&B	&08 50 07.7&$+36$ 04 30.3&\phantom{0}610	&$4.9\pm0.3$	&	1.8&&Unresolved\\
		&	&&&1400		&$1.10\pm0.06$	&	&&FIRST\\
		&C	&08 50 7.5&$+36$ 04 09.3&\phantom{0}610	&$18.8\pm1.0$	&1.1	&&Double source\\
		&	&&&1400		&$7.7\pm0.4$	&	&&FIRST\\\cmidrule{2-9}
		&Diffuse (D)&08 50 04.9 &$+36$ 04 28.6	&\phantom{0}610	&$5.2\pm0.6$	&	&210&Diffuse source\\
\noalign{\smallskip}
\hline\noalign{\smallskip}
{ A2552} &A	&23 11 33.3&$+03$ 38 06.0&\phantom{0}610	&$1.9\pm0.1$	&	&&BCG\\
		  &B	&23 11 31.0&$+03$ 37 55.7&\phantom{0}610	&$0.58\pm0.08$	&	&&\\
		  &C	&23 11 33.0&$+03$ 38 54.5&\phantom{0}610	&$0.99\pm0.09$	&	&&\\\cmidrule{2-9}
		  &D + C	&23 11 31.6&$+03$ 38 52.3	&\phantom{0}235	&$12.8\pm0.7$	&	&&\\
		  &D &23 11 31.6&$+03$ 38 52.3	&\phantom{0}610	&$8.5\pm0.5$	&	&180&\\		  
		  &Halo (H)&23 11 33.5&$+03$ 38 41.2&\phantom{0}610	&$\sim4.3$	&	&&Candidate\\
\noalign{\smallskip}
\hline\noalign{\smallskip}
{ Z3146}		&S1	&10 23 39.7&$+04$ 11 10.3&\phantom{0}610	&$7.0\pm0.4$	&$0.9$	&&BCG\\
	&	&&&1400	&3.3	&	&&G14\\
	&	&&&4900	&$1.42\pm0.07$	&$0.67$	&&G14\\	
	&	&&&8500	&$0.98\pm0.03$	&	&&G14\\\cmidrule{2-9}
	\noalign{\smallskip}
		&Mini-halo	&10 23 39.7&$+04$ 11 10.3&\phantom{0}610	&$11.5\pm0.6$	&	$0.9$&190&\\

	\noalign{\smallskip}
		&	&&&1400	&$5.2$	&$\sim 1$	&&G14\\
		\noalign{\smallskip}
		&	&&&4900	&$1.7\pm0.2$	&	&&G14\\
\noalign{\smallskip}
\hline\noalign{\smallskip}
{RXC\,J1504.1--0248}\dotfill	&S1	&15 04 07.5&$-02$ 48 15.1&\phantom{0}235	&$121\pm6$	&0.7&&$\alpha_{235}^{610}$\\
\noalign{\smallskip}
		&	&&&\phantom{0}327	&94	&	&&G11b\\
		\noalign{\smallskip}
		&	&&&\phantom{0}610	&$62.5\pm3.1$	&	&&\\
		\noalign{\smallskip}
		&	&&&1400	&$42\pm2$	&$0.56$&&$\alpha_{327}^{1400}$,G11b\\\cmidrule{2-9}
		\noalign{\smallskip}
		&Mini-halo	&15 04 07.5&$-02$ 48 15.1&235	&$87\pm4$\tablefootmark{f}	& &290&\\
		\noalign{\smallskip}
		&	&&&\phantom{0}327	&$121\pm6$	&1.3	&&$\alpha$ from Fig.~\ref{mhspec}\\
		\noalign{\smallskip}
		&	&&&\phantom{0}610	&$59\pm3$	&	&300&\\
		&	&&&1400	&$20\pm1$	&$1.24$&&$\alpha_{327}^{1400}$, G11b\\
		\noalign{\smallskip}
		\hline
\hline\noalign{\smallskip}		
\end{tabular}
\tablefoot{
\tablefoottext{a}{The spectral index ($\alpha$) is between the frequencies reported in two 
consecutive rows, unless stated in the comments.}
\tablefoottext{b}{May be contaminated by the extended components of A or C that are missed in FIRST.}
\tablefoottext{c}{VLA C-array data at 4.89 GHz (Giacintucci et al. in prep.).}
\tablefoottext{d}{Difference between the total flux density in low resolution image around S1 and that 
of S1 in high resolution image.}
\tablefoottext{e}{Difference between NVSS and FIRST flux densities.}
\tablefoottext{f}{May be affected by over subtraction of point source flux density.}
}
\end{table*}

\subsection{Non-detections and upper limits}

The method of injections to infer the detection limits on a radio halo flux density 
was introduced in the GRHS \citep[V08,][]{bru09} and is also continued to be used in the EGRHS (K13). 
A model radio halo of diameter 1 Mpc and a profile constructed from the profiles of 
observed radio halos \citep{bru07} is injected at the cluster center in the final self-calibrated visibilities 
using the AIPS task `UVMOD'. The resulting visibilities are imaged and the detection of the radio halo 
is evaluated. Model radio halos starting from high flux densities to gradually lower flux densities are injected 
until the detection in the image is similar to a marginal detection. 
The injected flux density corresponding to a ``hint'' of diffuse emission in these is considered as 
the upper limit for that particular data. The injections were carried out on 610 MHz data due to their higher 
sensitivities. 

\begin{figure*}
   \centering 
$
\begin{array}{cc}
\includegraphics[width=7 cm]{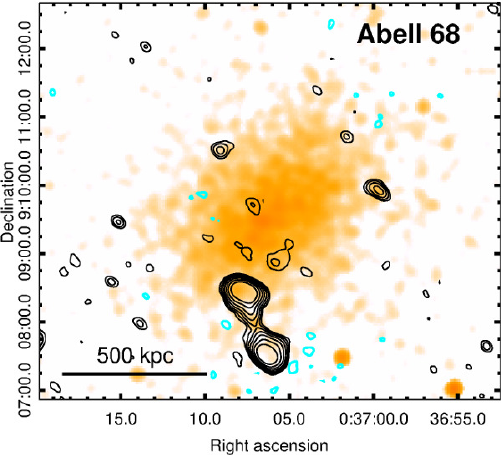}&
\includegraphics[width=7 cm]{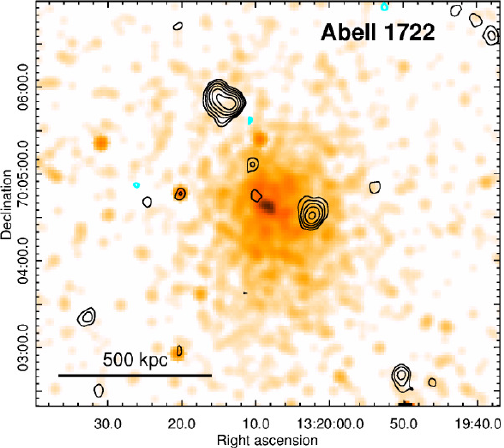}\\
\includegraphics[width=7 cm]{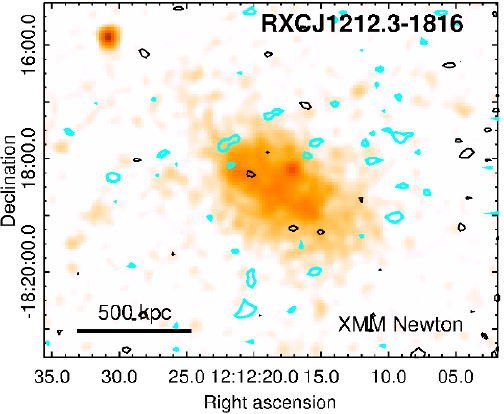}&
\includegraphics[width=7 cm]{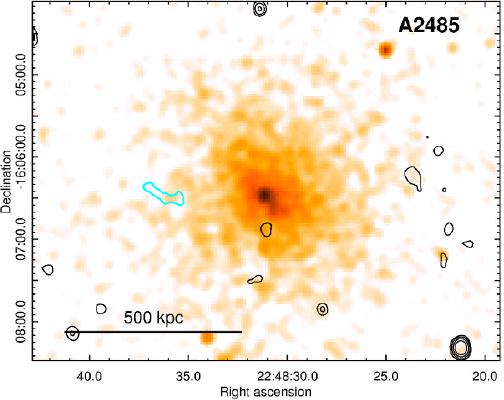}
\end{array}
$
      \caption{GMRT 610 MHz images of A68, A1722, RXCJ1212.3-1816 and A2485 
      in contours (black +ve and cyan -ve) 
      are shown overlaid on the respective Chandra X-ray images (colour). 
The contour levels are at $3\sigma\times(\pm1, 2, 4, 8, ...)$ in all the panels
(see Table 2 for $1\sigma$ levels and the beam sizes). 
The {0.5 - 2 keV exposure-corrected Chandra X-ray images with resolution $\sim4''$} are 
presented for A68 (Obs ID 03250), A1722 (Obs ID 03278) and A2485 (Obs ID 10439) 
and an XMM Newton (EPIC 
MOS) pipeline processed image ({Observation number 0652010201, 0.2 - 12 keV, resolution 
$\sim 8''$}) for RXCJ1212.3-1816 is presented. 
}
         \label{nondet}
\end{figure*}

The field of \object{A68} is dominated by a double lobed radio 
galaxy located at the southern edge of the cluster as seen in projection (Fig.~\ref{nondet}, upper left). 
There is no optical identification for this radio galaxy and thus it is likely a background source unrelated 
to the galaxy cluster. 
The cluster center is affected by the residuals around the radio galaxy that are likely to bias an upper limit 
based on a central injection. An upper limit of 6 mJy was obtained with injections in a region offset 
from the cluster center ($\sim 2'$ towards west) . 

A1722 is a bright, relaxed cluster as inferred from the X-ray properties. However 
the mass distribution is double peaked \citep{dah02}.
An upper limit of 3 mJy was obtained (Tab.~\ref{t4}). This is the deepest limit obtained among all the non-detections. 
In order to test the presence of a possible weak residual emission at the cluster center was biasing 
our upper limit lower, we 
also made injections offset from the cluster center ($\sim 5'$). Similar upper limits obtained in the offset 
injections enforces the significance of the upper limit in this cluster.

The cluster centers of RXCJ1212.3-1816 and A2485 (Fig.~\ref{nondet}) are void of radio sources.
In our earlier work (K13), upper limits based on 325 MHz data were presented for these clusters. 
Using the new 610 MHz data presented here, upper 
 limits of 6 and 5 mJy are estimated for these two clusters, respectively (Tab.~\ref{t4}). These are marginally 
 deeper than those reported in K13.

Here we also point out an important
 caveat regarding the interpretation of the upper limits.
 The upper limits are obtained using a model radio halo of a given size and 
 profile. The model is based on the profiles of a few well-known radio halos, however there is 
 a large variety in the morphologies of the observed radio halos as clear also from the radio halos 
 in this survey \citep[e.g. V07, V08][]{bru08}. 
 Therefore the upper limits 
 are to be treated in the context of the specific model and not as generalised upper limits on 
 the detection of diffuse emission with arbitrary or peculiar profiles.

In spite of such limitations, the upper limits are important outcomes of this survey and provide 
a tool to know where a cluster without a radio halo lies in the \lxlr plane relative 
to those with radio halos. 
Limits have a statistical meaning and are useful in population studies.
This method can be applied to any radio data to assess the detectability 
of a source of a particular morphology and brightness. 
The upcoming all sky surveys with the next generation instruments 
such as the LOFAR and MWA are expected to reach unprecedented surface brightness sensitivities and 
detect a large number of diffuse radio sources in galaxy clusters. 
The method used in EGRHS can be extended for use in these surveys to assess the non-detections.

\subsection{RXCJ0510.7-0801}
This cluster from the GRHS sample was observed at 610 and 235 MHz with the GMRT. 
The imaging of this cluster was affected by the presence of a strong radio source  
in the field of view. The rms noise achieved in the full resolution image at 610 MHz is 0.2 mJy beam$^{-1}$ 
and at 235 MHz is 1.2 mJy beam$^{-1}$ (Tab.~\ref{t2}). The images are presented in 
Fig.~\ref{apparxcj0510}. We have removed this cluster from 
the analysis due to the lack of more sensitive images.

\subsection{Imaging of known sources}

\subsubsection{Z3146}
\begin{figure*}
\centering
\includegraphics[height=4.8cm]{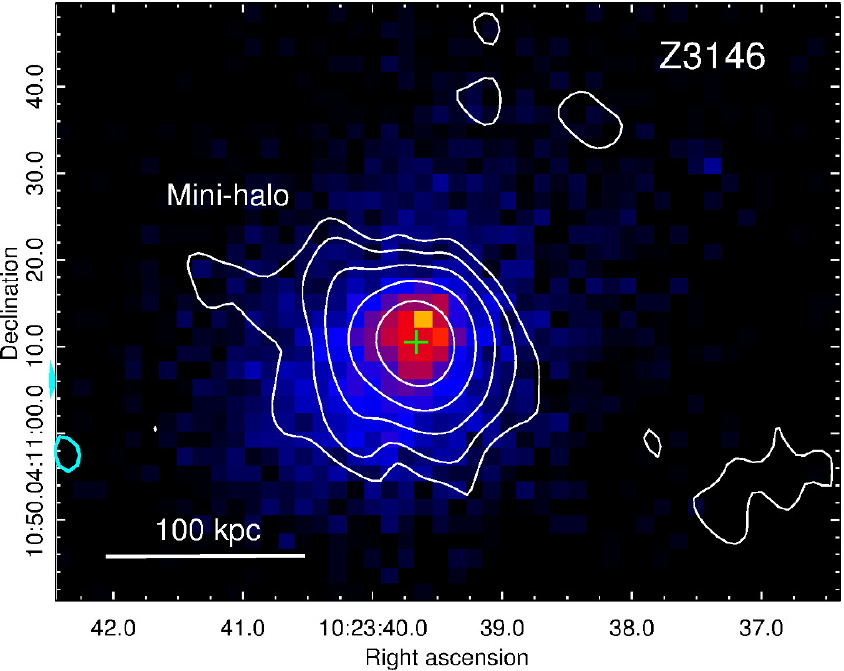}
\includegraphics[height=4.8cm]{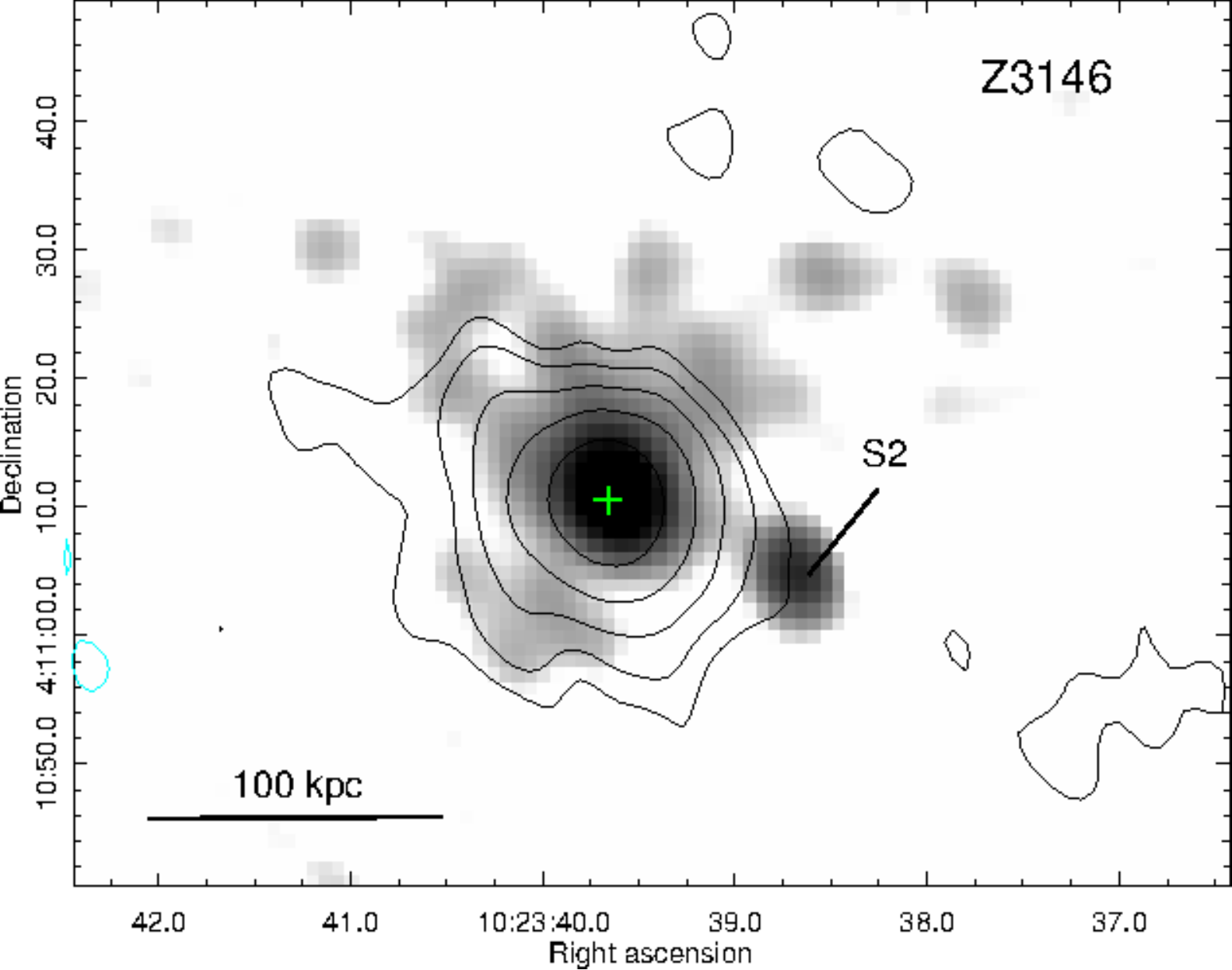}
\includegraphics[height=4.8cm]{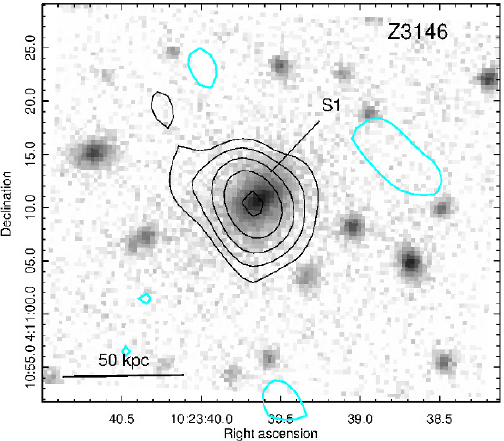}
\caption{{\bf Z3146}:{\it Left-} GMRT 610 MHz image in contours (white +ve and cyan -ve) 
overlaid on the Chandra X-ray image ({Obs ID 01651, 0.5 - 2 keV, resolution $\sim 2''$}) in colour.
The contours are at $0.3\times(-1, 1, 2, 4, ...)$ mJy beam$^{-1}$ and the beam is  
$8.9''\times7.1''$, p. a. $11.3^{\circ}$. 
 {\it Middle-} 
The 4.8 GHz image from G14 is shown in grey scale overlaid with the 610 MHz contours from 
the left panel.
{\it Right-} 
SDSS r-band image in colour showing the region near the BCG with
the high resolution 610 MHz image overlaid in contours (black +ve and cyan -ve). The 
 contours are at $0.3\times(\pm1, 2, 4, ...)$ mJy beam$^{-1}$ and the beam is 
$5.8''\times3.9''$, p. a. $33.8^{\circ}$.
}
\label{z3146}
\end{figure*}

Z3146 (\object{ZwCL\,1021.0+0426}) is a cool-core cluster in which a mini-halo was recently 
discovered using the 4.8 GHz VLA observations \citep[][hereafter, G14]{gia14}. 
We present the 610 MHz image of the cluster with the detection of the mini-halo.
The compact source S1 is associated with the central galaxy (Fig.~\ref{z3146}).
The flux density of S2 is expected to be below 3$\sigma$, consistent with our lack of detection.
The mini-halo surrounds the source S1 and has a largest extent of 190 kpc detected at 610 MHz (Fig.~\ref{z3146}). 
After subtracting the contribution from S1, the flux density of the mini-halo at 610 MHz is 11.5 mJy. 
This is consistent with the mini-halo spectral index $\sim 1$ over 1.4 - 4.8 GHz (G14) (Fig.~\ref{mhspec}). 
The different morphology of the mini-halo at 610 MHz and at 4.8 GHz (Fig.~\ref{z3146}, center) is due to 
the very different uv-coverage of the corresponding datasets, which span completely different hour angles.

\subsubsection{RXCJ1504.1-0248}

\begin{figure*}
\centering
\includegraphics[height=7cm]{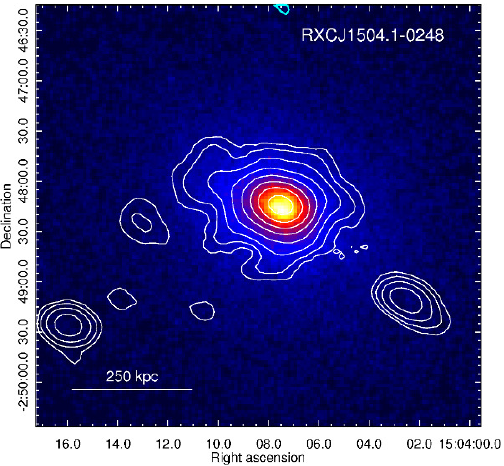}
\includegraphics[height=7cm]{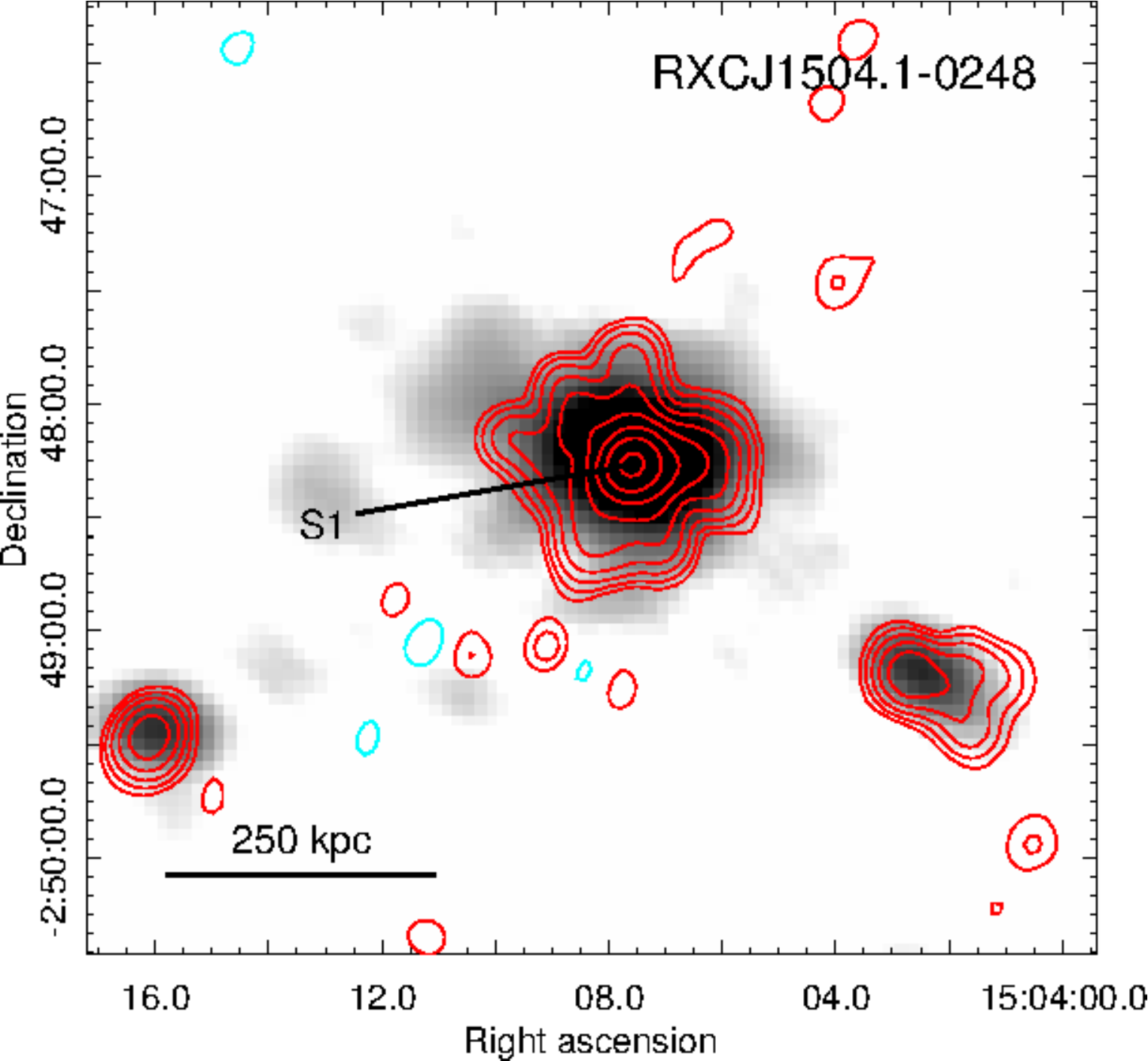}
\caption{{\bf RXC\,J1504.1-0248}:
{\it Left-} GMRT 610 MHz image in contours (white +ve and cyan -ve) 
overlaid on the Chandra X-ray image ({Obs ID 05793, 0.5 - 2 keV, resolution $\sim 2''$}) in colour.
The contours are at $0.4\times(-1, 1, 2, 4, ...)$ mJy beam$^{-1}$ and the beam is 
 $19.4''\times14.3''$, p. a. $70^{\circ}$.
{\it Right-} The 610 MHz image is shown in grey-scale with the 325 MHz image from G11b in contours 
(red +ve and cyan -ve). The 325 MHz contours are at 
$0.3\times(\pm1, 2, 4, ...)$ mJy beam$^{-1}$ and the beam is $11.3''\times10.4''$. 
}
\label{r1504}
\end{figure*}

\object{RXCJ1504.1-0248} is a cool-core cluster with a mini-halo detected at 327 MHz (GMRT) around its 
BCG \citep[][hereafter G11b]{gia11a}. We present the 610 and 235 MHz images of the mini-halo (Fig.~\ref{r1504}
 and Fig.~\ref{apparxcj1504}).
At 610 MHz the mini-halo has a largest linear extent of 290 kpc. At 325 MHz the mini-halo is roughly 
circular (radius $\sim 140$ kpc) in morphology and shows a sharp fall in the surface brightness at the edges 
(G11b). The morphology is more extended and smoother at 610 MHz (Fig.~\ref{r1504}). 
The spectral index of the mini-halo over the frequency range 235 -1400 MHz is estimated to 
be $\sim1.2$ (Fig.~\ref{mhspec}).

\section{Analysis of the full GRHS+EGRHS sample}\label{disc2}

The combination of EGRHS and GRHS is at present the largest uniformly radio surveyed sample of 
galaxy clusters to search for diffuse emission. In the following, we will refer to the GRHS+EGRHS 
as the full sample. The observations and the data  
 analysis are complete and the status of all the clusters in the full sample is reported (Tab.~\ref{tabonline}).
 The GRHS+EGRHS sample (64) consists of all the clusters in the redshift range 0.2 -- 0.4 that have X-ray luminosity, 
L$_{{\mathrm X}[0.1 -- 2.4 \mathrm{keV}]} > 5\times10^{44}$ erg s$^{-1}$ with $\delta > -31^{\circ}$. 
The distribution of the full sample in X-ray luminosity and redshift is shown in Fig.~\ref{lxz}. 

We can summarize the findings of diffuse emission (see Tab.~\ref{tabonline}, for references) in the full sample as follows :
\begin{itemize}
 \item 10 clusters have a radio halo (A209, A520, A697, A773, A1300, A1758a, A2163, A2219, RXCJ2003.5-2323, 
 RXCJ1514.9-1523);
 \item 2 clusters host a radio halo and a relic (A2744, A521);
 \item RXCJ1347.4-2515 has a radio halo and a double relic;
 \item 9 clusters host a mini-halo (RXCJ1504.1-0248, A1835, Z3146, RXJ1532.9+3021, S780, A3444
, A2390, RXJ2129.6+0005, Z7160).
\end{itemize}

Moreover, we found the following candidates:
\begin{itemize}
 \item 4 candidate radio halos (Z2661, A1682, Z5247, A2552);
 \item diffuse radio emission in Z1953, currently unclassified due to its size and location 
 in the cluster
 \item 3 candidate radio relics (A781, Z348, Z5247).
\end{itemize}

Upper limits on the flux density of a Mpc-sized radio halo were estimated 
for 30 clusters \footnote{The upper limit on RXJ2228.6+2037 is excluded.}. 
For three clusters with assessed absence of radio halo we were unable to set an upper limit
\footnote{A689- presence of a bright central source; A1763 and A2111 - VLA archival data 
at 1400 MHz, no 610 MHz data; A2146 - No new data, only literature information used.}.
Due to poor rms no upper limits were derived for A\,2813, A\,2895, A\,2645, A\,963 and RXC\,J0510.7-0801. 
Finally, no information is available for RXCJ2211.7-0350.
We removed these last six clusters from the statistical analysis and hence our 
statistical analysis is based on a total of 58 clusters.

\subsection{Fraction of radio halos, mini-halos and radio relics}

\begin{figure}
\centering
\includegraphics[height=8.2 cm,angle=0]{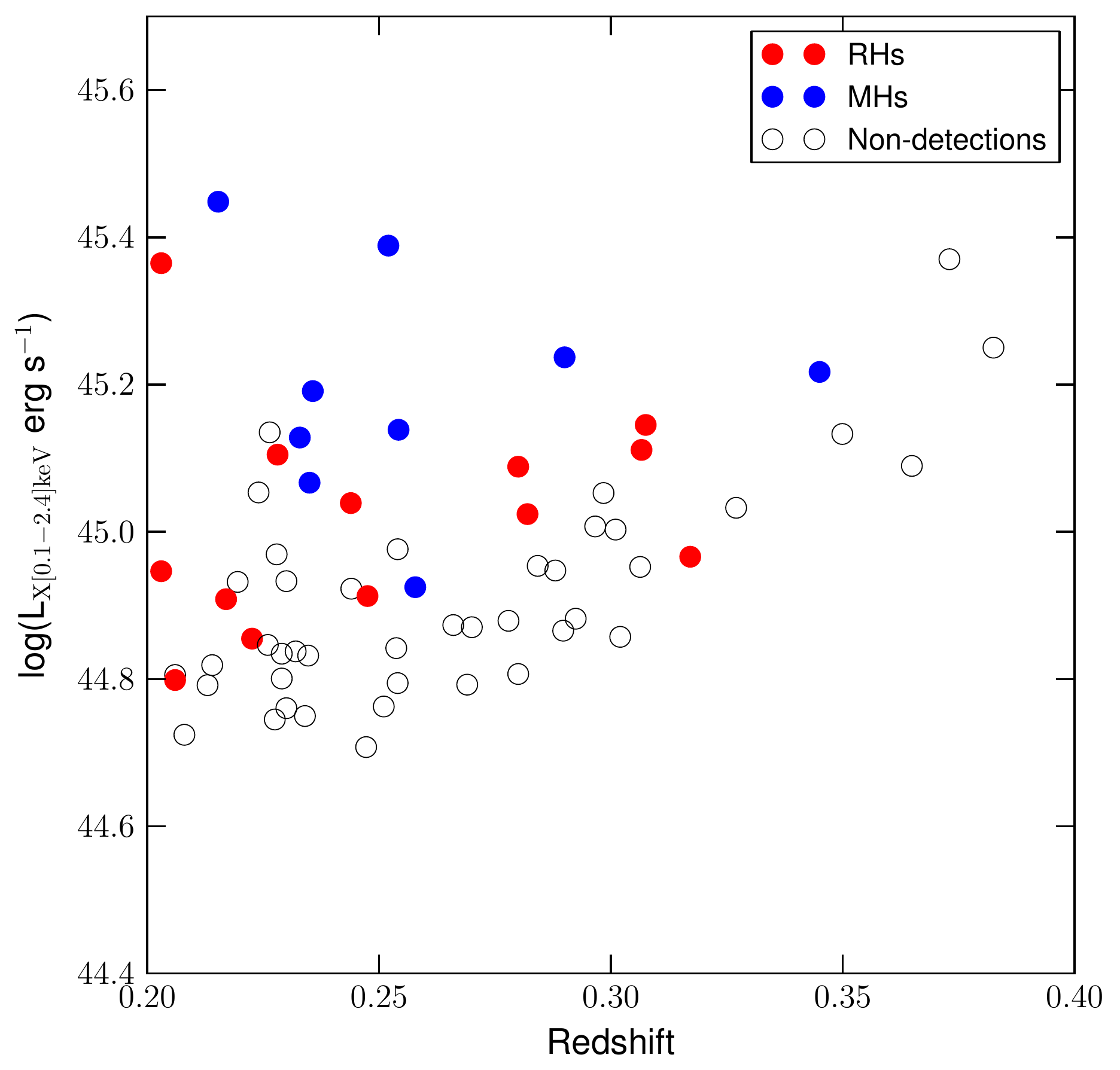}
\caption{The GRHS+EGRHS sample in the X-ray luminosity - redshift plane. 
}
\label{lxz}
\end{figure}

The fraction of radio halos in the full sample is 
$f_{\mathrm{RH}} = 13/58 =22\pm6 \%$  (here Poisson errors are used as a reference) and 
 $f_{\mathrm{RH}} = 17/58 = 29\pm7 \%$ if the 4 candidate radio halos are added.
The fraction of mini-halos is $f_{\mathrm{MH}} =$ $9/58 = 16\pm5 \%$. 
These fractions are specific to the considered samples and will change if 
the cluster selection is made from a different parent catalogue.

\begin{figure*}
\centering
\includegraphics[height=9 cm,angle=0]{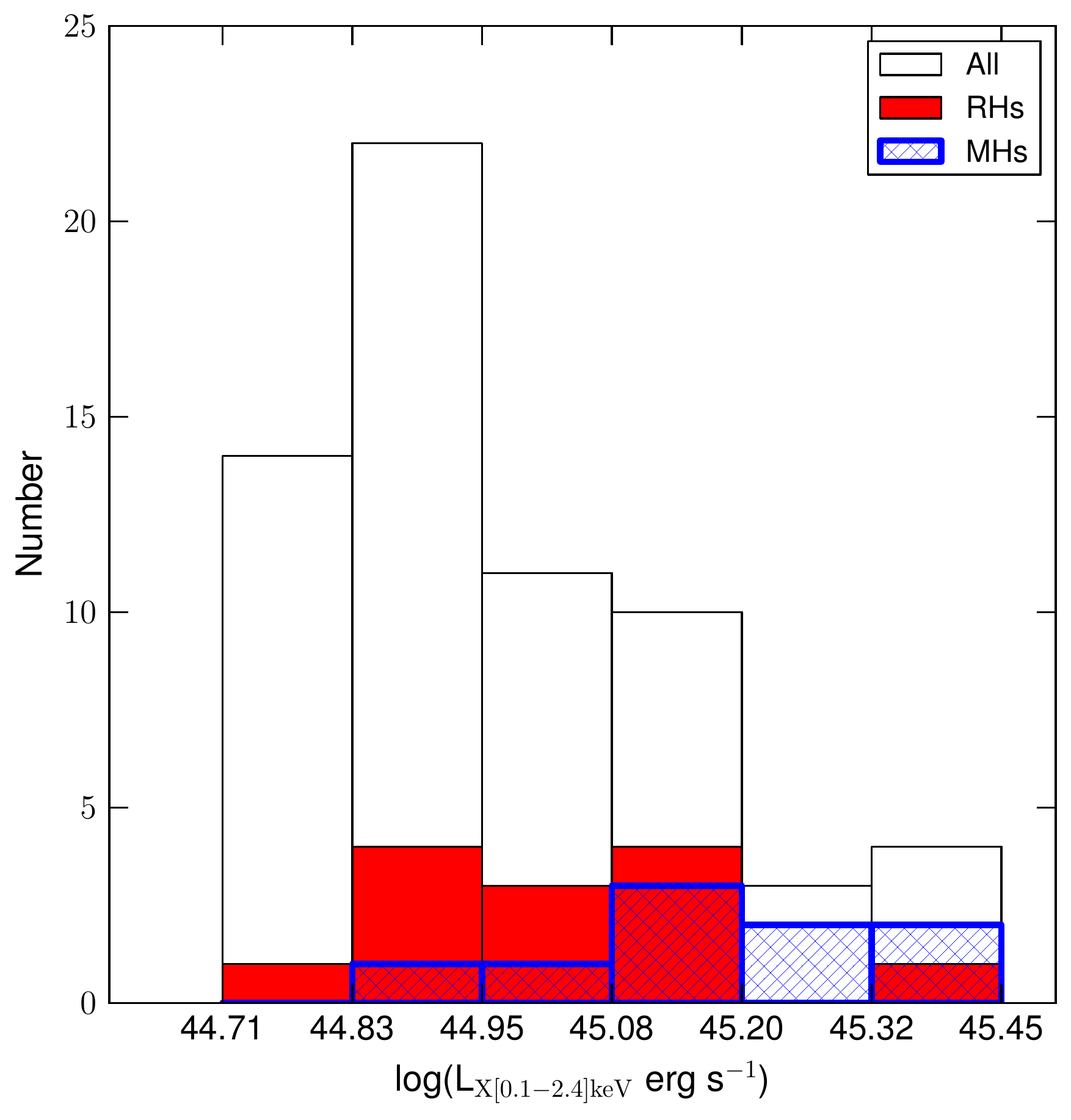}
\includegraphics[height=9 cm,angle=0]{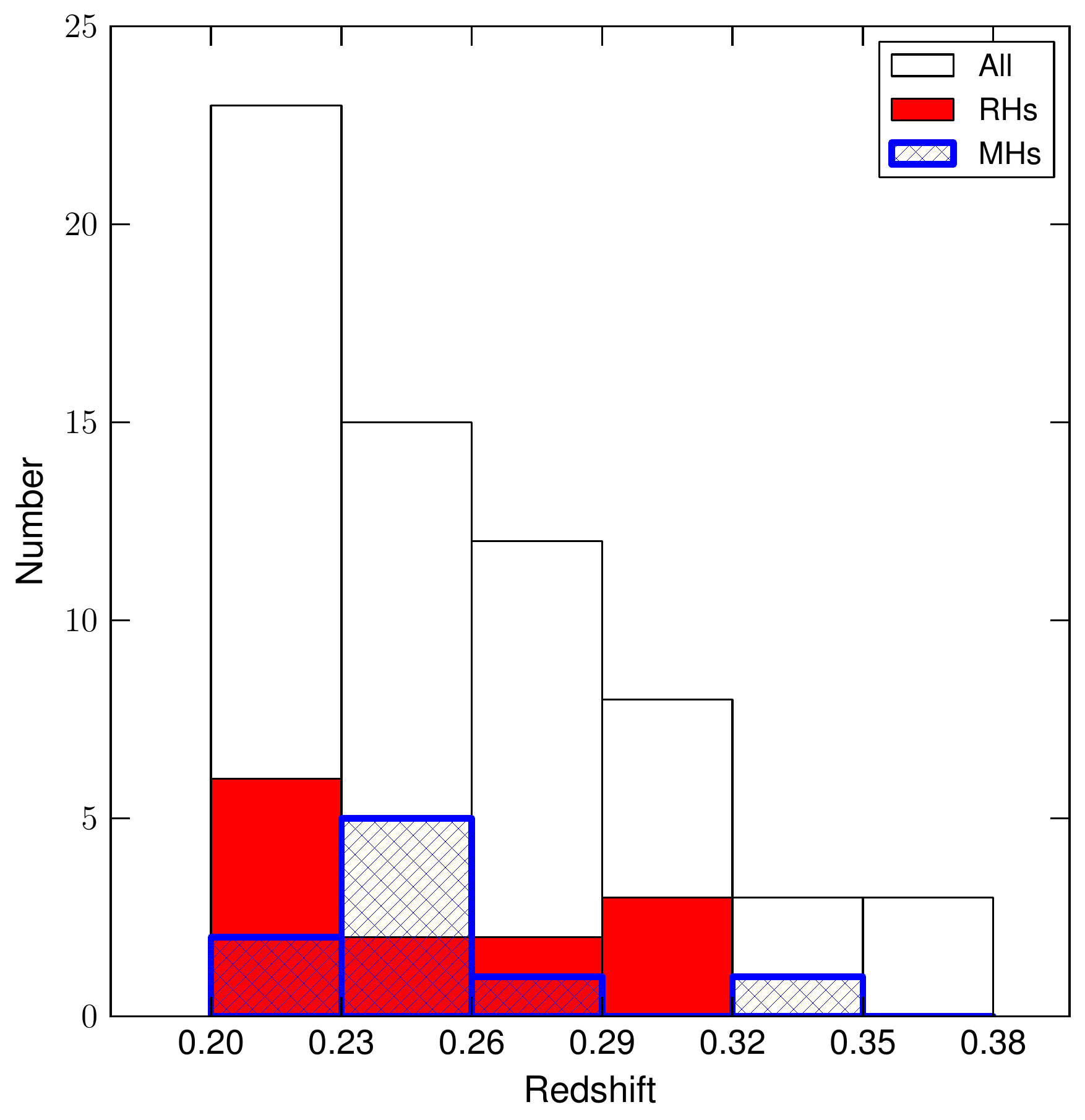}
\caption{Histrograms showing the distribution of the GRHS+EGRHS sample in 
X-ray luminosity (left) and in redshift (right). The white histogram 
includes all the clusters (All), 
the red includes those with radio halos (RHs) and the hatched includes those with 
mini-halos (MHs).
}
\label{hist}
\end{figure*}

The occurrence of radio halos and mini-halos as a function of X-ray luminosity and redshift is illustrated in 
the histograms (Fig.~\ref{hist}). It is striking from Fig.~\ref{lxz} and Fig.~\ref{hist} (left) 
that the mini-halos are detected in higher X-ray luminosity clusters in the sample; same is not the case with 
radio halos. 
  No statistically significant difference in the radio halo fraction in lower and higher luminosity 
  bins was found in our sample.
  However if the occurrence of radio halos and mini-halos is considered together, visual examination of 
  Fig.~\ref{hist} (left) indicates a higher occurrence for higher X-ray luminosity. We divided the sample 
into two bins using different values for the dividing X-ray luminosity and evaluated the occurrence of 
radio halos and mini-halos (RH+MH) in each bin. Using a Monte Carlo approach we found that the 
fraction, $f_{RH+MH}$ in the high X-ray luminosity bin differs with a $3.2\sigma$ significance 
from that in the low X-ray luminosity bin if the 
dividing X-ray luminosity is chosen to be $8\times10^{44}$ erg s$^{-1}$. In this case, the $f_{RH+MH}$ is $54\%$ 
in the high X-ray luminosity bin and $10\%$ in the low X-ray luminosity bin.
The very high X-ray luminosity 
clusters are typically either strong cool cores or very hot merging clusters and can explain the 
high occurrence of a MH or a RH. 

Our sample consists of 
X-ray luminous clusters selected from the flux limited X-ray catalogues of clusters. 
It has been speculated that with this method the relaxed, {cool-core} clusters are over represented, 
 leading to a lower estimate of the 
occurrence of radio halos 
as compared to a sample selected from a SZ selected cluster catalogue \citep{som14}; see \citet{cas13} 
for a first attempt to evaluate the occurrence of giant radio halos in a mass-selected sub-sample of 
GRHS+EGRHS (extracted from the Planck SZ catalogue).

Radio relics occur as single or double arc-like sources at the peripheries of clusters. 
In the GRHS+EGRHS sample, double radio relics are known in RXCJ1314.4-2515 and single radio relics are known in 
A521 and A2744. Candidate radio relics are detected in A781, Z348 and Z5247.
Thus, the fraction of radio relics is $f_{\mathrm{relics}} =$ $ 3/58 =$ $5\pm3 \%$. If 
the candidate relics are also included, the fraction is $10\pm 4\%$.

The overall occurrence of diffuse emission of any type (radio halos, mini-halos and relics) 
in the 58 clusters is $f_{\mathrm{diff}} =$ $25/58 =$ $43\pm9 \%$. 
Thus, in general the occurrence of diffuse emission in clusters does not seem 
rare in massive and X-ray luminous galaxy clusters. 
The all sky surveys with the next generation instruments such as the LOFAR and MWA 
with high sentivity to low brightness emission are expected reveal many more diffuse sources.

\subsection{The \lxlr \,diagram: halos and mini-halos}
\begin{figure}
\centering
\includegraphics[width=0.48\textwidth,angle=-90]{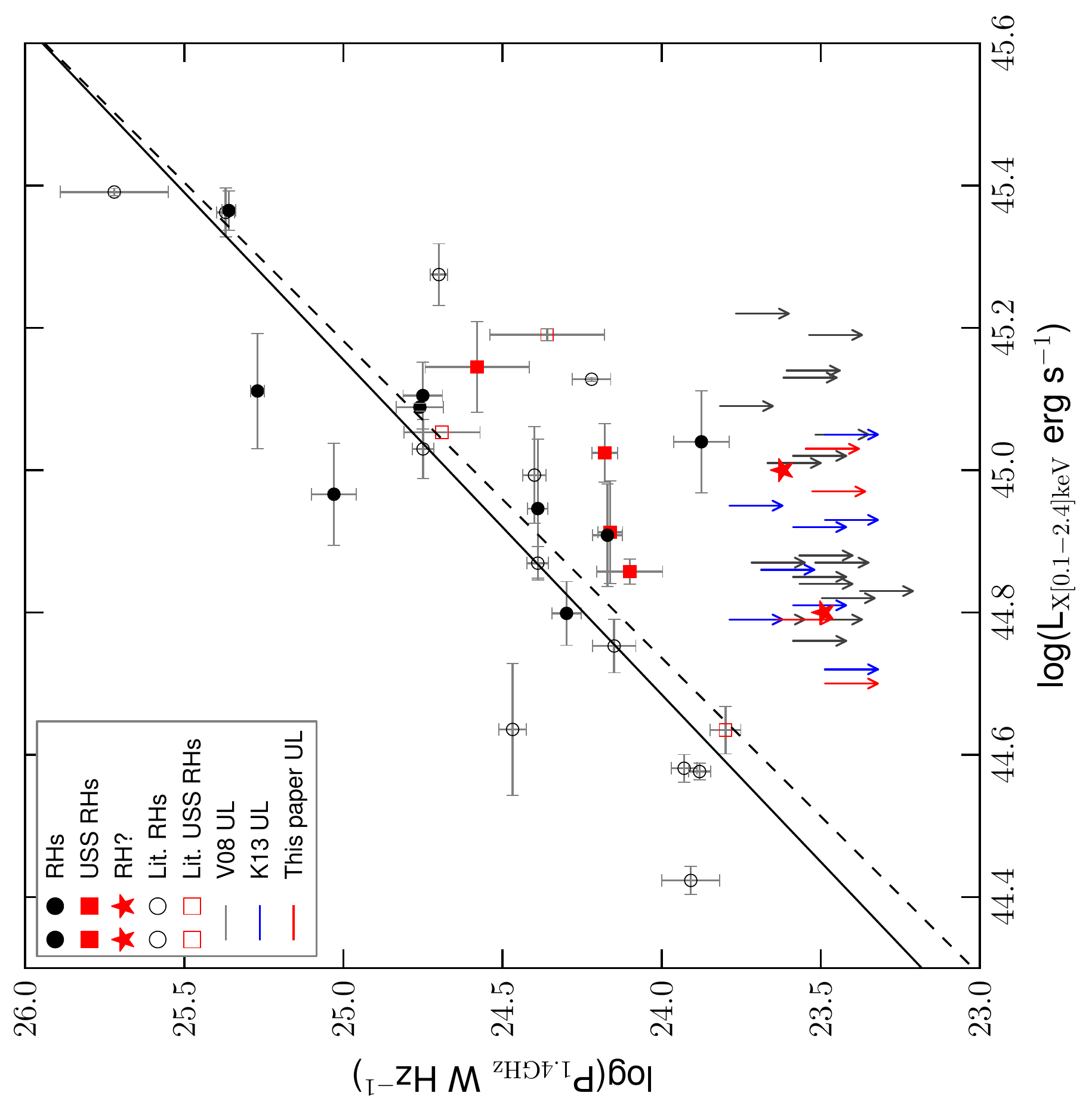}
\caption{The radio halos and the upper limits (UL) in the GRHS+EGRHS sample are shown in 
the \lxlr diagram along with the literature radio halos listed in \citet{cas13}. 
The filled symbols are for radio halos that 
belong to the GRHS+EGRHS sample and the empty symbols are for the literature 
radio halos. The squares are the USS radio halos. 
The arrows show the upper limits from the current (red) and previous works (grey, V08 and blue, K13). 
The best fit lines to the radio halos excluding the USS radio halos (solid) and including the USS 
radio halos (dashed) are also plotted.
}
\label{lxlrrh}
\end{figure}

One of the main results from the GRHS was the bimodal distribution 
of the radio-loud and radio-quiet (absence of a giant radio halo) galaxy clusters found in the \lxlr plane. The distribution in the 
\lxlr plane of the clusters surveyed in the GRHS+EGRHS is presented in Fig.~\ref{lxlrrh}.  The radio halos 
in literature listed in \citet{cas13} are also plotted.  
The sensitivity of the survey has allowed to clearly distinguish between the clusters with 
giant radio halos and those without. 
The ultra-steep spectrum (USS) (squares in Fig.~\ref{lxlrrh}) are radio halos with steep spectral indices 
($>1.5$) and are all below the empirical \lxlr correlation. 
The radio halo in RXC\,J1314.4-2515 is the weakest as compared to other radio halos in clusters with 
similar X-ray luminosity (the black point immediately above the upper limits in Fig.~\ref{lxlrrh}). On one hand we note that 
this is a giant radio halo, while on the other, it is possible that this source is still forming.

We also detected weak diffuse emission in two clusters, 
Z5247 and A2552 (candidate radio halos, Sec.~\ref{newdetect}) that falls in the region occupied by the upper limits 
(stars).
A linear fit using the BCES method \citep{akr96} of the form,
\begin{equation}
 log(P_{\mathrm{1.4 GHz}}) = A \times log(\mathrm{L}_{X}) + B
\end{equation}
was carried out for the radio halos. 
We carried out separate fits by including and excluding the USS halos in the data.
The best fit parameters are: $A=2.13\pm0.26$ and $B=-70.97\pm11.52$ (solid line in 
Fig.~\ref{lxlrrh}) and $A=2.24\pm0.28$ and $B=-76.41\pm12.65$ (dashed line in 
Fig.~\ref{lxlrrh}), excluding and including, respectively, the USS radio halos from the data. 

\begin{figure}
\centering
\includegraphics[width=0.48\textwidth,angle=-90]{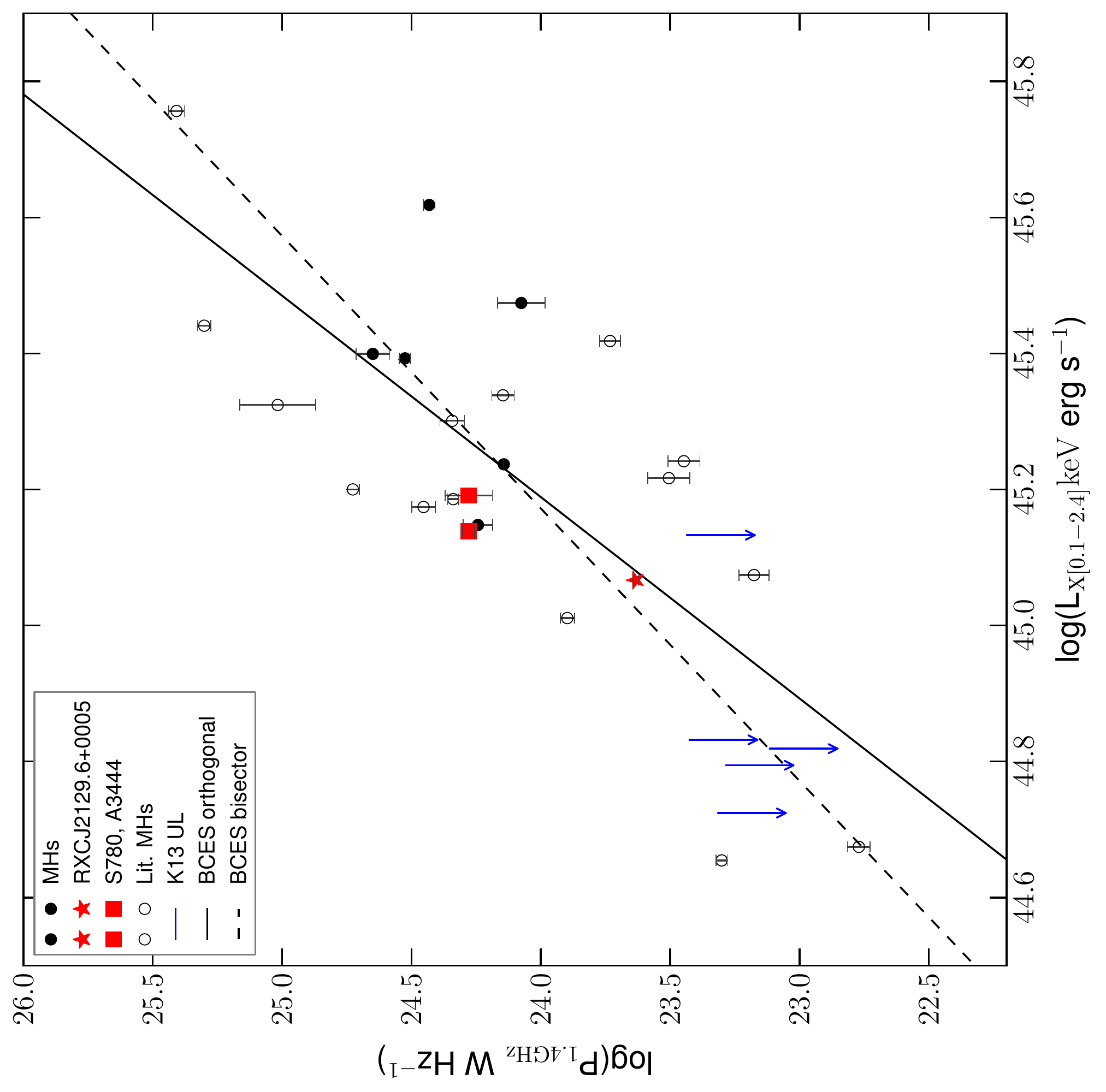}
\caption{\label{lxlrmh}The mini-halos (filled symbols) and the upper limits (UL) from the GRHS+EGRHS are shown in 
the \lxlr \,diagram along with the literature mini-halos (empty symbols). The best fit lines using the 
BCES bisector (dashed) and orthogonal (solid) methods are also plotted.
}

\end{figure}

The distribution of clusters with mini-halos in the \lxlr diagram is 
reported in Fig.~\ref{lxlrmh}, together with the upper-limits for mini-halo detections in cool cores 
(K13).
We found a possible trend 
between $P_{1.4\,GHz}$ and the cluster X-ray luminosity. We fit this relation with 
Eq. 1 obtaining the best-fit parameters: $A = 2. 49\pm0. 30$, $B = - 88. 74\pm13. 92$ 
(dashed line in Fig.~\ref{lxlrmh}) and $A =  3. 37\pm 0. 70$, $B = - 128. 60\pm32. 07$ 
(solid line in Fig.~\ref{lxlrmh}), by using the the bisector and orthogonal methods, respectively.
There is a large scatter in the distribution and the two methods lead to different slopes for the 
scaling.
The upper limits obtained for the mini-halos (K13) are conservative and are not enough for 
testing if there is bimodality.
There are two main reasons for the conservative upper limits. The central bright 
radio source poses difficulties in obtaining deeper upper limits. Injections 
offset from the center will be deeper, however will not represent the real mini-halos that are found 
superposed on the central radio sources. In addition, the model radio halo profile scaled to the 
size of 500 kpc was used as a model for mini-halo injection (K13). The observed profiles of mini-halos are 
more centrally peaked than that of the radio halos \citep{mur09} and can 
lead to slightly upper limits.

\subsection{Cluster dynamics in GRHS+EGRHS}\label{clusdyn}
The non-thermal emission detected in the forms of radio halos, relics and mini-halos are 
linked to the various forms of disturbances in the ICM.
There is strong observational evidence to support the hypothesis that dynamical disturbance in clusters 
is closely connected to the generation mechanism of radio halos \citep[e.g.][]{cas10}.
Radio relics at cluster outskirts are tracers of shocks in cluster mergers.
In the case of mini-halos, the turbulence injected by the sloshing cool-core may  
play a role \citep[e.g.][]{git06,gia14,zuh13}.
The connection between the dynamical state and the diffuse emission needs to be closely examined 
making use of a large sample like GRHS+EGRHS.

\begin{figure*}
\centering
 \includegraphics[height=7 cm]{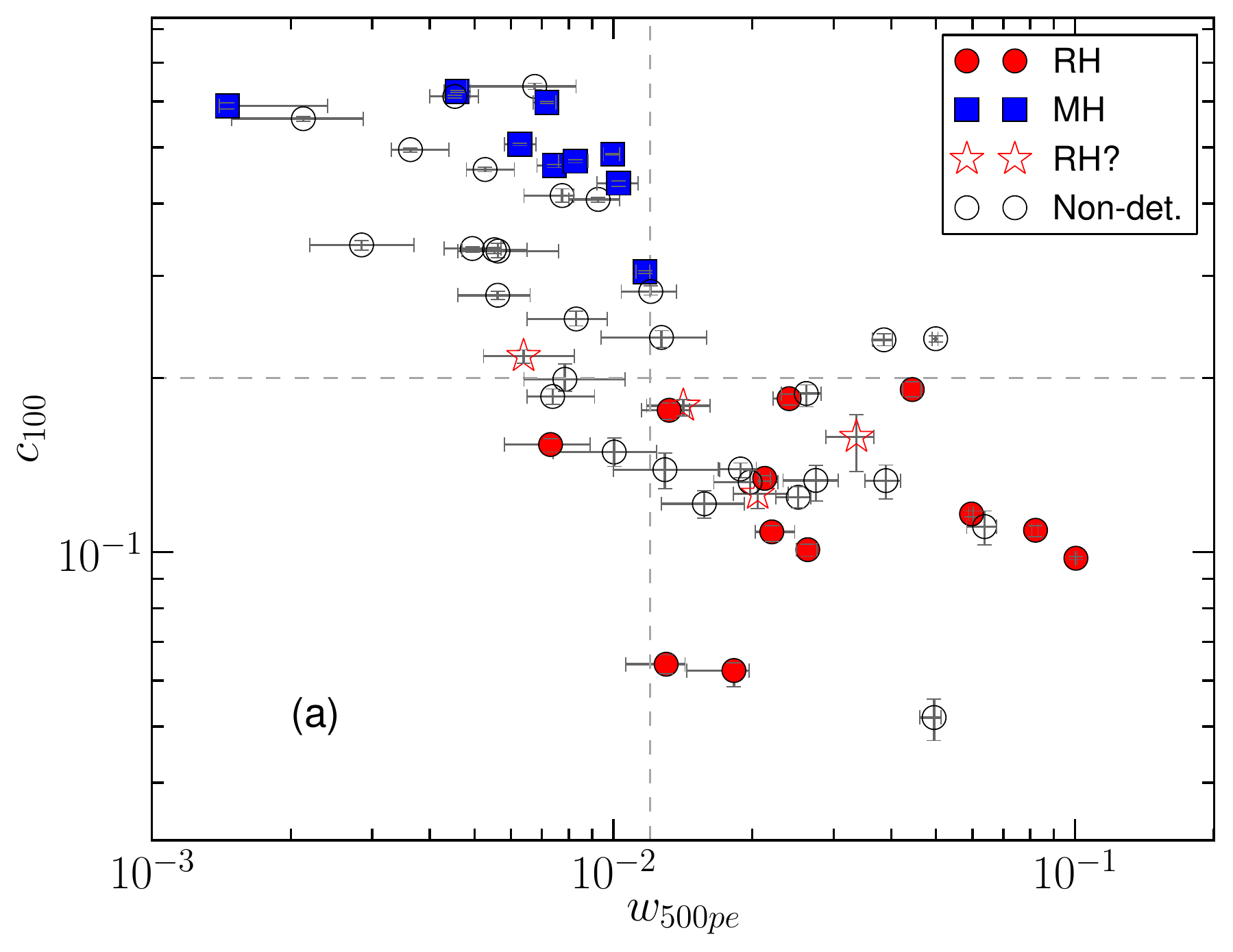}
 \includegraphics[height=7 cm]{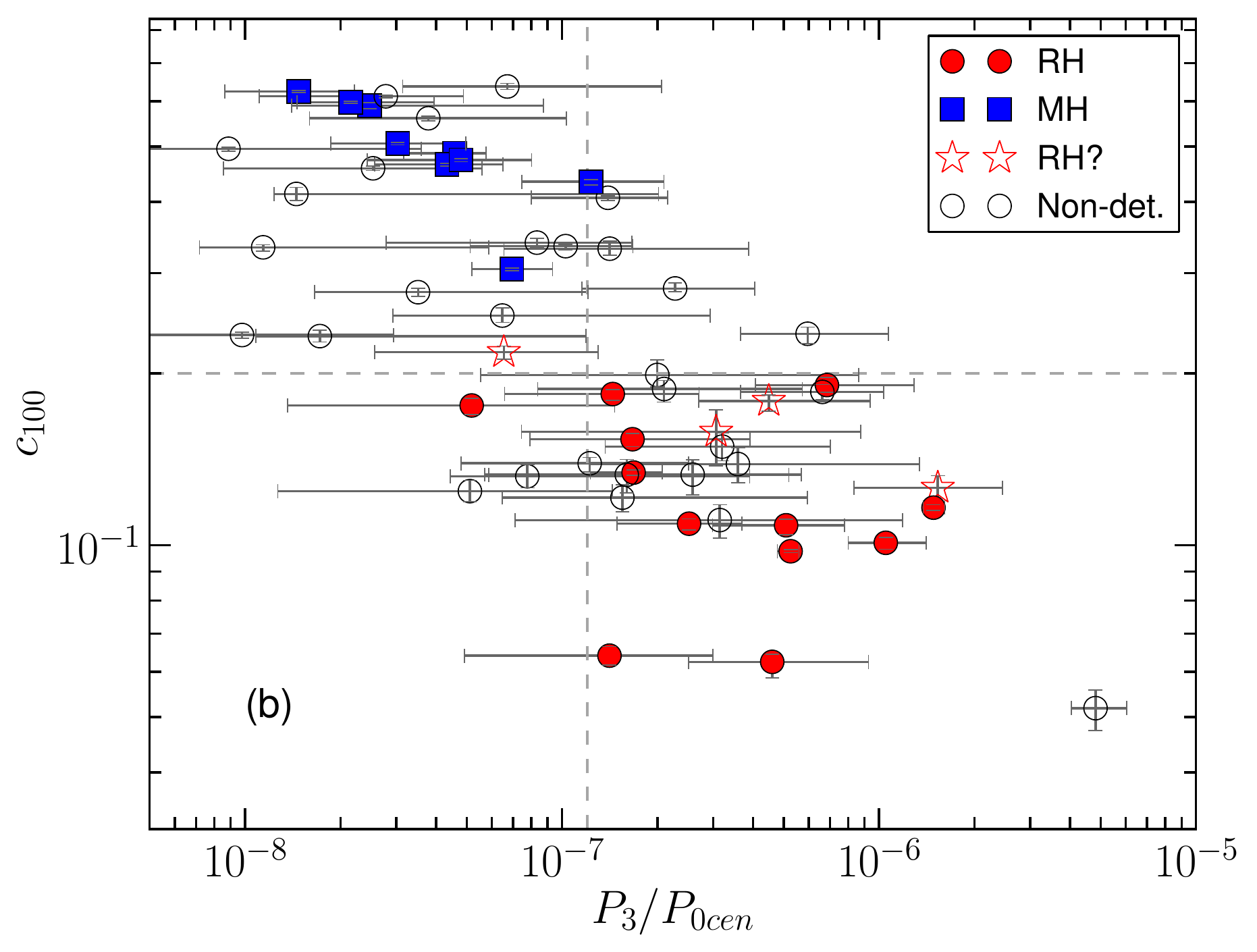}\\
 \includegraphics[height=7 cm]{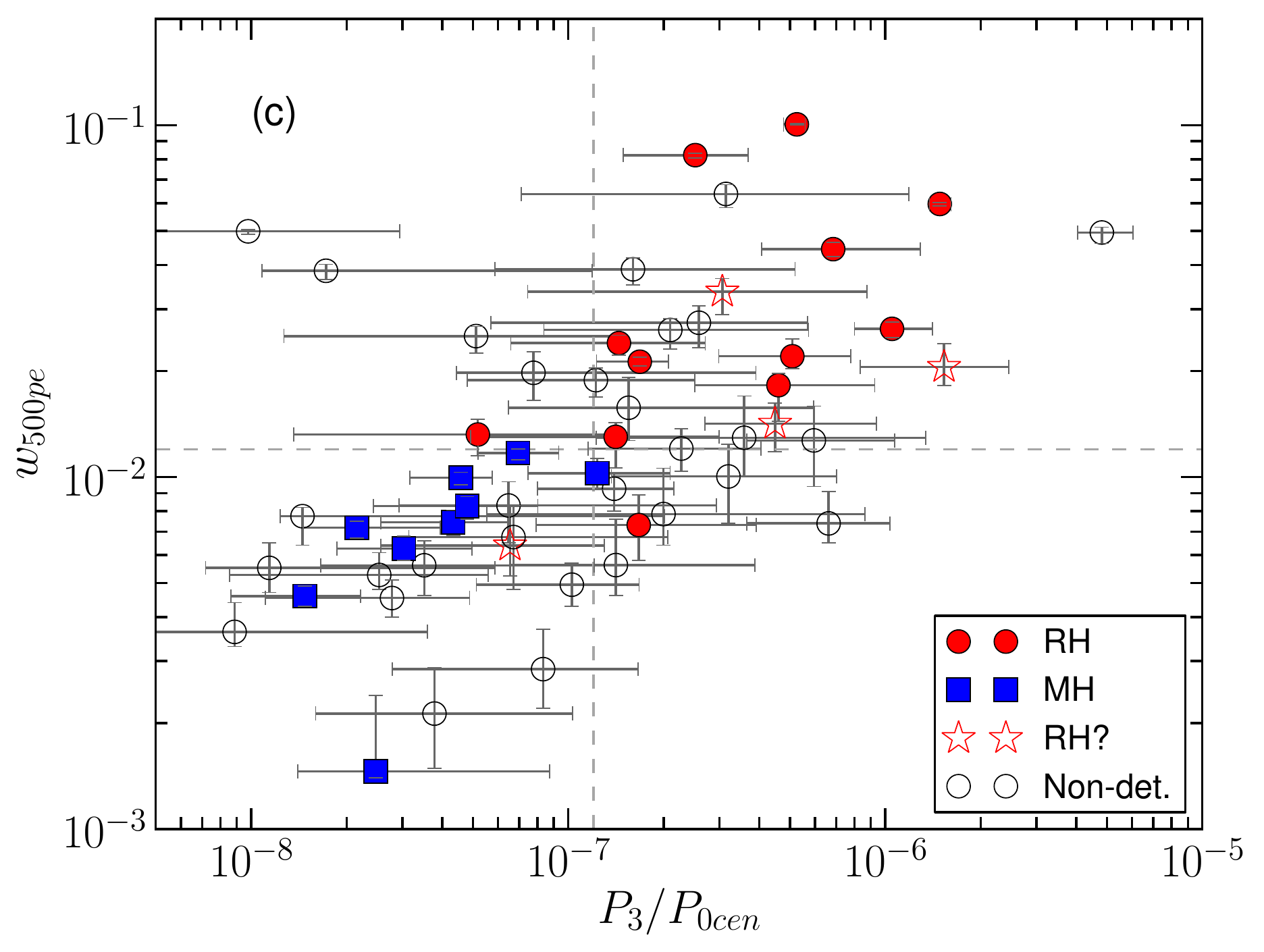}
\caption{The GRHS+EGRHS clusters are shown in the planes formed by the morphological parameters, namely,
the concentration parameter ($c_{100}$), the centroid shift ($w_{500pe}$) and the power ratio 
($P_3/P_0$). The clusters that are non-detections (empty 
circles), those with radio halos (red filled circles) and those with mini-halos (blue squares) and 
those with candidate radio halos (A1682, Z5247, A2552 and Z2661; red stars) 
are marked. The grey dashed lines mark 
the boundaries proposed by \citet{cas10} between the merging and the non-merging clusters for each of the parameters. 
}
\label{morph}
\end{figure*}

The X-ray surface brightness maps represent the state of the thermal ICM which 
is collisional and thus reveal the dynamical state of a cluster.
The disturbance in the ICM can be quantified using the morphological state estimators 
on the X-ray surface brightness maps \citep[e.g.][]{cas10}. 
We used the cluster morphological state estimators described in \citep{cas10}, 
namely, the concentration parameter ($c_{100}$), the centroid shift ($w_{500pe}$) and the power ratios 
($P_{3}/P_{0cen}$) available for the GRHS+EGRHS clusters  \citep[][Cuciti et al. in prep.]{cas10,cas13}.
The sample clusters are shown in the planes formed by these parameters with 
the markers distinguishing the clusters with radio halos, mini-halos and with no radio emission 
(Fig.~\ref{morph}).
  { While the clusters with mini-halos and radio halos appear well separated into 
quadrants for less and more disturbed clusters, 
the clusters which do not have any radio emission populate the region of both mini-halos and
 radio halos (Fig.~\ref{morph}).}
 
 The fact that several merging clusters do not host a radio halo provides crucial information 
 for the origin of these sources.
If turbulence plays a role in the acceleration and dynamics 
 of relativistic particles radio halos evolve in a time-scale of about 1 Gyr. According to 
 this scenario, the possibility to generate a halo that is sufficiently luminous (detectable 
 by our observations) at the GMRT frequencies depends on the combination of several parameters, 
 including the magnetic field properties of the hosting cluster, the mass of the merging systems, 
 the fraction of the energy of mergers that is drained in turbulence (and its properties) and 
 the stage of the {merger \citep[][for a review]{cas10,rus11,don13, bru14}.
 Further statistical} studies that consider also information on 
 the mass of clusters will be very useful.
Interestingly those merging clusters that do not 
 show halos at the level of current upper limits are also candidate to host ultra-steep-spectrum 
 halos that should be visible at lower frequencies \citep{bru08,2010A&A...517A..10C};
  in this respect future LOFAR observations will be particularly useful.

\section{Summary and conclusions}\label{sum}
The GRHS and EGRHS together are a sample of 64 bright galaxy clusters 
surveyed at low frequency bands (610/235 MHz) to understand the statistical properties of diffuse 
radio emission in galaxy 
clusters. The sample consisted of all the galaxy clusters  
in the redshift range 0.2 -- 0.4 that have X-ray luminosities, 
$L_{X[0.1 - 2.4\mathrm{keV}]} > 5 \times 10^{44}$ erg s$^{-1}$ and are above 
the declination of $-31^{\circ}$ selected from the REFLEX and the eBCS catalogues. 
This selection was tailored to suit the sensitivity of 
the GMRT at 610 MHz and also matched the redshift in which most of the radio halos 
are expected to occur based on the turbulent reacceleration models. 
The results from the GRHS, based on the radio information on 35 clusters 
were presented (V07, V08), after which the extension of the sample, called the EGRHS was carried 
out. The first part of the EGRHS data were presented in K13. In this work we 
have presented the final batch of 11 galaxy clusters and the statistical results based on the 
full sample. 

 Radio images of 11 galaxy clusters at 610 
 and 235 MHz with rms noise in the ranges 0.03 -- 0.2 mJy beam$^{-1}$ and 0.5 -- 1.2 mJy beam$^{-1}$, 
 respectively, are presented and discussed. 
  We discovered: a mini-halo in the cluster RXC\,J2129.6+0005, a relic and a candidate radio halo 
  in Z5247, a candidate radio halo in A2552, and an unclassified diffuse source in Z1953.
 The detections of mini-halos in the 
 GRHS clusters S780 and A3444 are also reported. The 610 MHz images
 of the mini-halos in Z3146 and RXC\,J1504.1-0248 are compared with those in literature. 
 Combining the 610 and 235 MHz data with that available from the literature, 
 the integrated spectra of the mini-halos in the clusters Z3146, RXCJ1504.1-0248 and RXC\,J2129.6+0005 are presented. 
 The spectral indices are in the range 0.9 - 1.3.
 The method of model radio halo injection was used on 4 clusters in this work.

On the completion of the presentation of radio data analysis for all the clusters we 
have carried out a statistical analysis of the full sample.
The results are as follows:
\begin{enumerate}
 \item There are a total of 64 clusters in the sample of which 48 were part of the GRHS and 16 additional 
 formed the EGRHS sample. Those which had insufficient radio data in GRHS (V07, V08) were also included for 
 new dual frequency (610-235 MHz) GMRT observations with the EGRHS. The full sample can 
 be found in the Online Table.~\ref{tabonline}.
 \item The primary aim of the survey was to search for radio halos, though other forms of diffuse emission were 
recorded. There are 13 radio halos in the sample; 5 were discovered in the GRHS 
 and 8 were known from literature. The fraction of radio halos in the full sample is $f_{\mathrm{RH}} \sim 22\%$ 
 (and $\sim29\%$ including the candidates). 
\item There are 9 mini-halos in the sample of which new radio data on 5 mini-halos were presented in 
this survey. The fraction of mini-halos in the sample is $f_{\mathrm{MH}} \sim 16\%$.
\item The combined occurrence fraction of radio halos and mini-halos is $\sim54\%$ in the clusters with 
higher X-ray luminosities ($L_X > 8\times10^{44}$ erg s$^{-1}$) as compared to 
that of $\sim10\%$ in the lower X-ray luminosities. Based on Monte Carlo analysis this drop is significant at 
$3.2\sigma$ level.
 \item The fraction of clusters with confirmed radio relics in our sample is $f_{\mathrm{relics}} \sim 5 \%$. 
 If the candidate relics were included, the fraction is $\sim 10\%$. The relics are thus found to be even 
 rarer than the radio halos in this sample. 
 \item 
 In the cases of non-detection of a radio halo, an upper limit was determined.
 A total of 31 upper limits for the detections of radio halos with 1 Mpc size and of 
 a model profile based on observed radio halos are reported in this survey. The upper limits 
 are a unique outcome of this survey and provide vital grounding to the occurrence fractions 
 of radio halos reported in this work.
 \item The \lxlr diagram for the radio halos including all the available information on detections and upper limits 
 from the survey is presented. The bimodality in the distribution of detections and upper limits is evident. 
 \item The mini-halos are also plotted in the \lxlr diagram along with the upper limits reported in K13. A bimodality is not 
 seen in the distribution, though it cannot be ruled out. The mini-halo upper 
 limits were based on a model radio-halo of size 500 kpc and thus are
 conservative. 
 \item 
 The morphological status estimators, namely, the concentration parameter, centroid shift and power ratios evaluated 
  using the archival Chandra data for the survey clusters were used as proxies 
  for the dynamical state of the cluster. The regions occupied by the radio halos and mini-halos 
 in the planes formed by these parameters are distinct; the radio halos occur in merging clusters whereas mini-halos in 
 relaxed clusters. The clusters with non-detections are found in the merging as well as non-merging quadrants.
\end{enumerate}
The completion of this survey provides the first uniformly surveyed sample on the occurrence of 
radio halos. 
The methods for analysis presented in this survey can be applied in a broader context of future surveys 
with the next generation telescopes such as the JVLA, LOFAR, MWA and eventually the SKA.

\begin{table}
\caption{\label{t4} Radio power upper limits using radio halo injections. 
{Spectral index of 1.3 was used to scale the flux density to 1.4 GHz.}}
\centering
\begin{tabular}{lclll}
\hline
Cluster & rms$^a$&S$_{610\mathrm{MHz}}$ &Log ($P_{1.4\; \mathrm{GHz}}$\\
Name  & mJy b$^{-1}$&mJy& W Hz$^{-1}$ ) \\
\hline \noalign{\smallskip}
RXC\,J1212.3-1816  &0.08&6& 23.65 \\
A2485&0.11& 5& 23.50\\
A1722 & 0.06 &3 &23.56\\ 
A68 &0.05 &6$^{b}$ & 23.54\\
\hline
\end{tabular}
\tablefoot{$^a$ Rms noise in images tapered to HPBW $\sim 18''-25''$.\\
$^{b}$ An injection $2'$ away from the center to avoid a bright radio galaxy.}
\end{table}

\begin{acknowledgements}
We thank the staff of the GMRT who have made these observations possible. 
GMRT is run by the National Centre for Radio Astrophysics of the Tata Institute
of Fundamental Research. The National Radio Astronomy Observatory is a facility 
of the National Science Foundation operated under cooperative agreement by 
Associated Universities, Inc.
This research made use of the NASA/IPAC
Extragalactic Database (NED) which is operated by the Jet Propulsion Laboratory,
California Institute of Technology, under contract with the National Aeronautics
and Space Administration. We made use of the ROSAT Data Archive of the
Max-Planck-Institut fur extraterrestrische Physik (MPE) at Garching, Germany.
This research made use of data obtained from the High Energy Astrophysics
Science Archive Research Center (HEASARC), provided by NASA's Goddard Space
Flight Center. This work is partially supported by PRIN-INAF2008 and by
FP-7-PEOPLE-2009-IRSES CAFEGroups project grant agreement 247653. GM 
acknowledges financial support by the ``Agence Nationale de la 
Recherche'' through grant ANR-09-JCJC-0001-01. RK thanks J. Donnert 
for insightful discussions.
\end{acknowledgements}
\bibliographystyle{aa}
\bibliography{kale_egrhs_arxiv}

\Online

\begin{appendix} 

\onecolumn
\section{Online Table}
\begin{longtable}{llllllll}
\caption{\label{tabonline} GRHS+EGRHS sample and results. The columns are: 1. Cluster name; 2. 
Right Ascension (J2000); 3. Declination (J2000); 4. redshift; 5. X-ray luminosity; 
6. 1.4 GHz radio power of the radio halo, mini-halo or upper limit. {Spectral 
index of 1.3 was used to scale the values where the power was not available at 1.4 GHz.}
7. {Remark: H= Halo, Re= Relic, USS = Ultra-steep spectrum, mH = mini-halo, 
DRe = Double relic,} UL = upper limits, (c) = candidate, 
mhUL = mini-halo upper limit, Brt. Src. =Bright source at the center and thus UL not estimated, 
bad rms = UL not estimated 
due to poor rms noise in the map; 8. References for $P_{1.4\mathrm{GHz}}$: (V07) \citet{ven07}; 
(V08) \citet{ven08}; 
(K13) \citet{kal13}; (K14) This paper; 
(G11a) \citet{gia11a}; (G11b) \citet{gia11b}; (B09) \citet{bru09}; (F12) \citet{fer12};
(R12) \citet{rus12}; (G in prep.)  Giacintucci et al. in prep.; (C13) \citep{cas13}.}\\
\hline\noalign{\smallskip}
Name	  	& RA$_{J2000}$ & DEC$_{J2000}$ & $z$ & $L_{X[0.1-2.4 \mathrm{keV}]}$ & 
log$_{10} (P_{1.4\mathrm{GHz}}$ & Remark & Ref.\\
\noalign{\smallskip}
		& hh mm ss  & $^{\circ}$ $'$ $''$   &   & $10^{44}$ erg s$^{-1}$ & 
		$ \mathrm{W Hz}^{-1} $)  & &\\
\hline
\noalign{\smallskip}
\endfirsthead
\caption{continued.}\\
\hline\noalign{\smallskip}
Name	  	& RA$_{J2000}$ & DEC$_{J2000}$ & $z$ & $L_{X[0.1 - 2.4 \mathrm{keV}]}$ & 
log$_{10} (P_{1.4\mathrm{GHz}}$ & Remark & Ref.\\
\noalign{\smallskip}
		& hh mm ss  & $^{\circ}$ $'$ $''$   &   & $10^{44}$ erg s$^{-1}$ & 
		$ \mathrm{W Hz}^{-1} $)  & &\\
\noalign{\smallskip}
\hline
\noalign{\smallskip}
\endhead
\hline
\endfoot
\endlastfoot
A2697 	  	& 00  03  11.8 & -06  05  10 & 0.232 & 6.88 & $<23.60$  & UL & V08 \\
A2744 	  	& 00  14  18.8 & -30  23  00 & 0.307 & 12.92 & $25.27\pm0.02$ & H+Re & B09 \\
A2813 	  	& 00  43  24.4 & -20  37  17 & 0.292 & 7.62 & -- &  bad rms & V08 \\
A68 	  	& 00  36  59.4 & +09  08  30 & 0.254 & 9.47 & $<23.54$ &  UL(off-center) & K14 \\
RXJ0027.6+2616 	& 00  27  49.8 & +26  16  26 & 0.365 & 12.29 & $<23.83$ & UL & V08 \\
A141 		& 01  05  34.8 & -24  39  17 & 0.23 & 5.76 & $<23.60$  & UL & V08 \\
A209 		& 01  31  53.0 & -13  36  34 & 0.206 & 6.29 & $24.07\pm0.09$ & H & V08, B09 \\
A267 		& 01  52  52.2 & +01  02  46 & 0.23 & 8.57 & $<23.50$  & UL & K13 \\
A2895 		& 01  18  11.1 & -26  58  23 & 0.228 & 5.56 & -- & bad rms & V08 \\
RXJ0142.0+2131	& 01  42  3.1 & +21  30  39 & 0.28 & 6.41 & $<23.62$  & UL & K13 \\
Z348 (RXCJ0106.8+0103)	& 01  06  50.3 & +01  03  17 & 0.254 & 6.23 & $<23.78$  & UL, Re (c) & K13 \\
		& 	       &             &       &      & $<23.30$  & mhUL & K13 \\
A3088 		& 03  07  4.1 & -28  40  14 & 0.254 & 6.95 & $<23.60$  & UL & V08 \\
A520 		& 04  54  19.0 & +02  56  49 & 0.203 & 8.84 & $24.59\pm0.04$ & H & B09 \\
A521 		& 04  54  9.1 & -10  14  19 & 0.248 & 8.18 & $24.06\pm0.04$ & USS H+Re & B09 \\
RXCJ0437.1+0043	& 04  37  10.1 & +00  43  38 & 0.284 & 8.99 & $<23.78$  & UL & K13 \\
RXJ0439.0+0520 	& 04  39  2.2 & +05  20  43 & 0.208 & 5.3 & $<23.47$  & UL & K13 \\
		& 	       &             &       &      & $<23.33$  & mhUL & K13 \\
RXJ0439.0+0715 	& 04  39  1.2 & +07  15  36 & 0.244 & 8.37 & $<23.63$  & UL & K13 \\
RXCJ0510.7-0801	& 05  10  44.7 & -08  01  06 & 0.220 & 8.55 & -- & bad rms & K14 \\
A611 		& 08  00  58.1 & +36  04  41 & 0.288 & 8.86 & $<23.60$  & UL & V08 \\
A689		& 08  37  29.7 & +14  59  29 & 0.279 & 2.8  & -- & Brt src. & K13 \\
A697 		& 08  42  53.3 & +36  20  12 & 0.282 & 10.57 & $24.10\pm0.09$ & USS H & V08 \\
Z1953 (ZwCl0847.2+3617)	& 08  50  10.1 & +36  05  09 & 0.373 & 23.46 & -- & (c) & K14 \\
A773 		& 09  17  59.4 & +51  42  23 & 0.217 & 8.1 & $24.23\pm0.04$ & H & B09 \\
A781 		& 09  20  23.2 & +30  26  15 & 0.298 & 11.29 & $<23.53$ & UL, Re (c) & V07, V08 \\
Z2089 (ZwCl0857.9+2107)	& 09  00  45.9 & +20  55  13 & 0.235 & 6.79 & $<23.39$  & UL & V08 \\
		& 	       &             &       &      & $<23.44$  & mhUL & K13 \\
Z2661 (RXCJ0949.8+1707)	& 09  49  57.0 & +17  08  58 & 0.383 & 17.79 & -- & H (c) & V08 \\
Z2701 (ZwCl0949.6+5207)	& 09  52  55.3 & +51  52  52 & 0.214 & 6.59 & $<23.51$  & UL & V08 \\
		& 	       &             &       &      & $<23.13$  & mhUL & K13 \\
A3444 		& 10  23  50.8 & -27  15  31 & 0.254 & 13.76 & $24.27\pm0.02$  & mH & G (in prep), K14 \\
A963 		& 10  17  9.6 & +39  01  00 & 0.206 & 6.39 & --  & bad rms & V07,V08 \\
Z3146 (ZwCl1021.0+0426)	& 10  23  39.6 & +04  11  10 & 0.29 & 17.26 & $24.13$ & mH& G14 \\
A1300 		& 11  31  56.3 & -19  55  37 & 0.308 & 13.97 & $24.78\pm0.04$ & USS H& B09  \\
A1423 		& 11  57  22.5 & +33  39  18 & 0.213 & 6.19 & $<23.55$  & UL & V08 \\
RXCJ1115.8+0129	& 11  15  54.0 & +01  29  44 & 0.350 & 13.58 & $<23.63$ & UL & V08 \\
		& 	       &             &       &      & $<23.45$  & mhUL & K13 \\
A1576 		& 12  36  49.1 & +63  11  30 & 0.302 & 7.2 & $<23.78$  & UL & K13 \\
RXCJ1212.3-1816	& 12  12  18.9 & -18  16  43 & 0.269 & 6.2 & $<23.65$  & UL & V08, K13 \\
Z5247 (RXCJ1234.2+0947)	& 12  34  17.3 & +09  46  12 & 0.229 & 6.32 & $23.49$ & H (c) & K14 \\
A1682 		& 13  06  49.7 & +46  32  59 & 0.226 & 7.02 & $24.34$ & H (c) & V08 \\
A1722 		& 13  19  43.0 & +70  06  17 & 0.327 & 10.78 & $<23.56$ & UL & K13, K14 \\
A1758a 		& 13  32  32.1 & +50  30  37 & 0.28 & 12.26 & $24.75\pm0.17$ & H & C13	 \\
A1763 		& 13  35  17.2 & +40  59  58 & 0.228 & 9.32 & -- & No H & V07, V08 \\
RXCJ1314.4-2515	& 13  14  28.0 & -25  15  41 & 0.244 & 10.94 & $23.87\pm0.08$ & DRe+H & V08 \\
Z5699 (ZwCl1303.6+2647)& 13  06  0.4 & +26  30  58  &0.306 & 8.96 & $<23.73$  & UL & V08 \\
Z5768 (ZwCl1309.1+2216)	& 13  11  31.5 & +22  00  05 & 0.266 & 7.47 & $<23.53$ & UL & V08 \\
A1835 		& 14  01  2.0 & +02  51  32 & 0.252 & 24.48 & $24.05$ & mH & F12 \\
S780 		& 14  59  29.3 & -18  11  13 & 0.236 & 15.53 & $<23.55$  & UL & V08 \\
		& 	       &             &       &      & $24.27\pm0.1$  &  mH & G (in prep), K14 \\
Z7160 (ZwCl1454.8+2233)& 14  57  15.2 & +22  20  30 & 0.258 & 8.41 & $24.24\pm0.06$ & mH & V08, G14 \\
A2111 		& 15  39  38.3 & +34  24  21 & 0.229 & 6.83 & -- & No H & V08 \\
A2146 		& 15  56  4.7 & +66  20  24 & 0.234 & 5.62 & -- & No H & R12 \\
RXCJ1504.1-0248	& 15  04  7.7 & -02  48  18 & 0.215 & 28.08 & $24.47$ & mH & G11b \\
RXCJ1514.9-1523	& 15  14  58.0 & -15  23  10 & 0.223 & 7.16 & $24.23$ & USS H& G11a, C13 \\
RXJ1532.9+3021	& 15  32  54.2 & +30  21  11 & 0.362 & 16.49 & $24.52\pm0.02$ & mH & K13, G14 \\
Z7215 (	RXCJ1501.3+4220)& 15  01  23.2 & +42  21  06 & 0.290 & 7.34 & $<23.73$ & UL & V08 \\
A2163 		& 16  15  46.9 & -06  08  45 & 0.203 & 23.17 & $25.26\pm0.005$ & H& B09 \\
A2219 		& 16  40  21.1 & +46  41  16 & 0.228 & 12.73 & $25.08\pm0.02$ & H& B09 \\
A2261 		& 17  22  28.3 & +32  09  13 & 0.224 & 11.31 & $<23.48$ & UL & V08, K13 \\
RXCJ2003.5-2323	& 20  03  30.4 & -23  23  05 & 0.317 & 9.25 & $25.08\pm0.03$ & H & B09 \\
A2390 		& 21  53  34.6 & +17  40  11 & 0.233 & 13.43 & $24.98\pm0.02$ & mH & B09\\
RXJ2129.6+0005 	& 21  29  37.9 & +00  05  39 & 0.235 & 11.66 & $23.69$ & mH & K14 \\
A2485 		& 22  48  32.9 & -16  06  23 & 0.247 & 5.1 & $<23.50$ & UL & K13, K14 \\
RXCJ2211.7-0350	& 22  11  43.4 & -03  50  07 & 0.27 & 7.42 & --  & -- & -- \\
RXJ2228.6+2037 	& 22  28  34.4 & +20  36  47 & 0.418 & 19.44 & $<23.93$ & UL & V08 \\
A2537 		& 23  08  23.2 & -02  11  31 & 0.297 & 10.17 & $<23.68$ & UL & V08 \\
A2552 		& 23  11  33.1 & +03  38  37 & 0.305 & 10.07 & -- & H (c) & K14 \\
A2631 		& 23  37  40.6 & +00  16  36 & 0.278 & 7.57 & $<23.58$ & UL & V08 \\
A2645 		& 23  41  16.8 & -09  01  39 & 0.251 & 5.79 & -- & bad rms & V08 \\
A2667 		& 23  51  40.7 & -26  05  01 & 0.226 & 13.65 & $<23.62$ & UL & V08 \\
\hline
\end{longtable}

\twocolumn

\section{Radio images of the cluster fields}\label{appfig}
The radio images of the clusters up to the virial radius are presented here.

\begin{figure*}
\centering
\includegraphics[width=9cm]{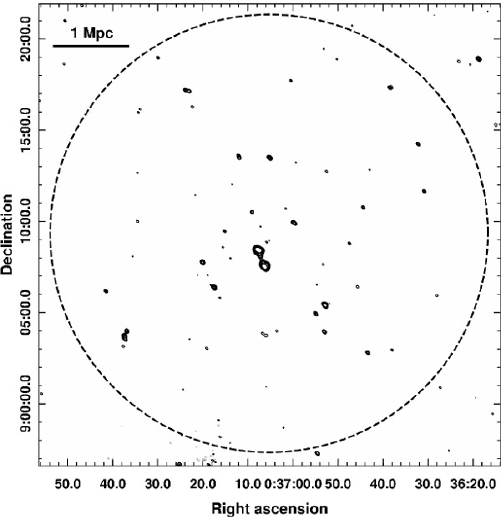}
\includegraphics[width=9cm]{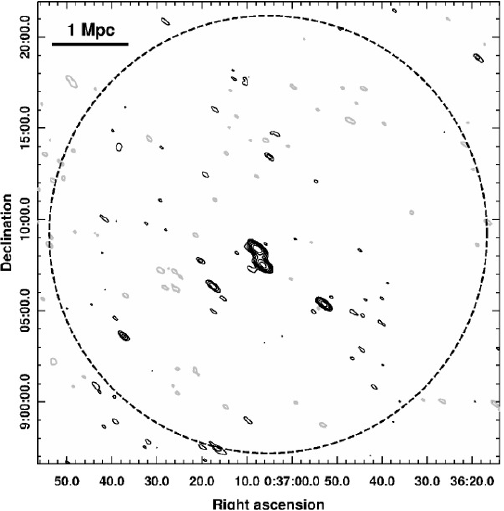}
\caption{{\bf A68}: GMRT 610 MHz (left) and 235 MHz (right) images in 
contours. Positive
contours are black, and negative ones are grey in all the images in Appendix A. 
The contours are at $0.33\times(\pm1, 2, 4,...)$ mJy beam$^{-1}$ at 610 MHz 
and $6.0\times(\pm1, 2, 4,...)$ mJy beam$^{-1}$ at 235 MHz.  The
HPBWs at 610 and 235 MHz are $9.3''\times6.5''$, p. a. $47.9^{\circ}$ and 
$29.0''\times12.7''$, p. a. $52.2^{\circ}$, respectively. The 
$1\sigma$ levels in the 610 and 235 MHz images are 0.05 and 
1.5 mJy beam$^{-1}$, respectively. The circle has a radius equal to the virial 
radius for this cluster ($12.0'$).}
\label{appa68}
\end{figure*}

\begin{figure*}
\centering
\includegraphics[width=9cm]{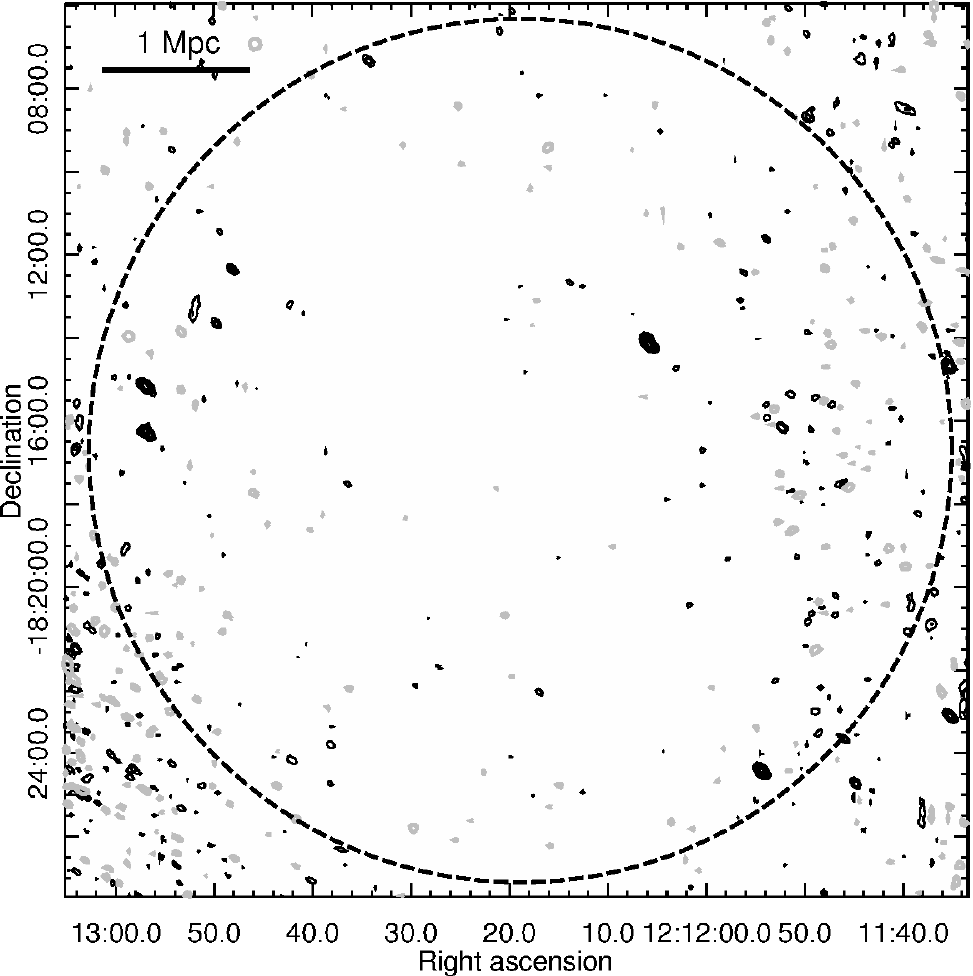}
\caption{{\bf RXCJ1212.3-1816}: GMRT 610 MHz image in 
contours.
The contours are at $0.3\times(\pm1, 2, 4,...)$ mJy beam$^{-1}$. The
HPBW at 610 is $11.4''\times5.7''$, p. a. $47.6^{\circ}$. The 
$1\sigma$ levels in the 610 MHz is 0.07 mJy beam$^{-1}$. The circle has a radius equal to the virial 
radius for this cluster ($10.4'$).}
\label{appar1212}
\end{figure*}

\begin{figure*}
\centering
\includegraphics[width=9cm]{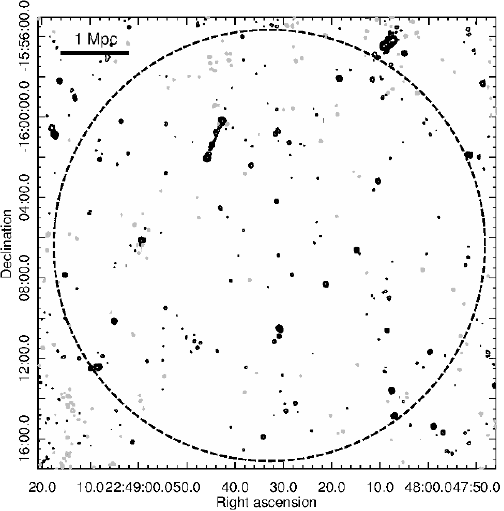}
\includegraphics[width=9cm]{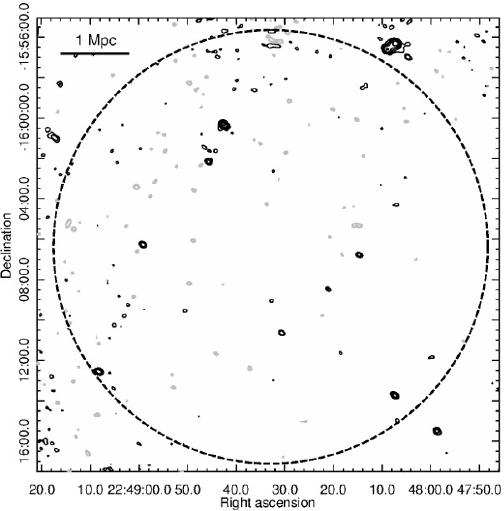}
\caption{{\bf A2485}: GMRT 610 MHz (left) and 235 MHz (right) images in 
contours. 
The contours are at $0.18\times(\pm1, 2, 4,...)$ mJy beam$^{-1}$ at 610 MHz 
and $3.0\times(\pm1, 2, 4,...)$ mJy beam$^{-1}$ at 235 MHz.  The
HPBWs at 610 and 235 MHz are   $8.1''\times7.0''$, p. a. $9.2^{\circ}$
 and  $14.9''\times11.5''$, p. a. $50.3^{\circ}$, respectively. The 
$1\sigma$ levels in the 610 and 235 MHz images are 0.06 and 
1.0 mJy beam$^{-1}$, respectively. The circle has a radius equal to the virial 
radius for this cluster ($10.7'$).}
\label{appa2485}
\end{figure*}

\begin{figure*}
\centering
\includegraphics[width=9cm]{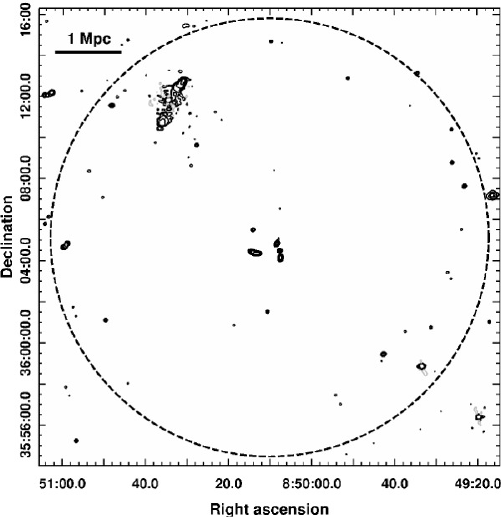}
\includegraphics[width=9cm]{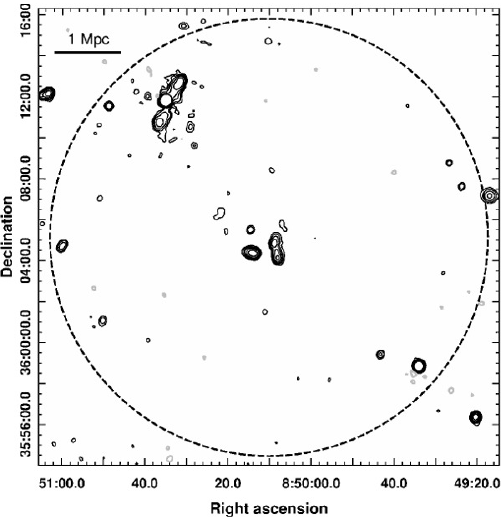}
\caption{{\bf Z1953 \,{\bf (ZwCl0847.2+3617)}}: GMRT 610 MHz (left) and 235 MHz (right) images in 
contours. 
The contours are at $0.45\times(\pm1, 2, 4,...)$ mJy beam$^{-1}$ at 610 MHz 
and $1.8\times(\pm1, 2, 4,...)$ mJy beam$^{-1}$ at 235 MHz.  The
HPBWs at 610 and 235 MHz are $7.2''\times6.9''$, p. a. $-57.8^{\circ}$
 and  $17.8''\times15.9''$, p. a. $-14.7^{\circ}$, respectively. The 
$1\sigma$ levels in the 610 and 235 MHz images are 0.06 and 
0.6 mJy beam$^{-1}$, respectively. The circle has a radius equal to the virial 
radius for this cluster ($10.6'$).}
\label{appaz1953}
\end{figure*}



\begin{figure*}
\centering
\includegraphics[width=9cm]{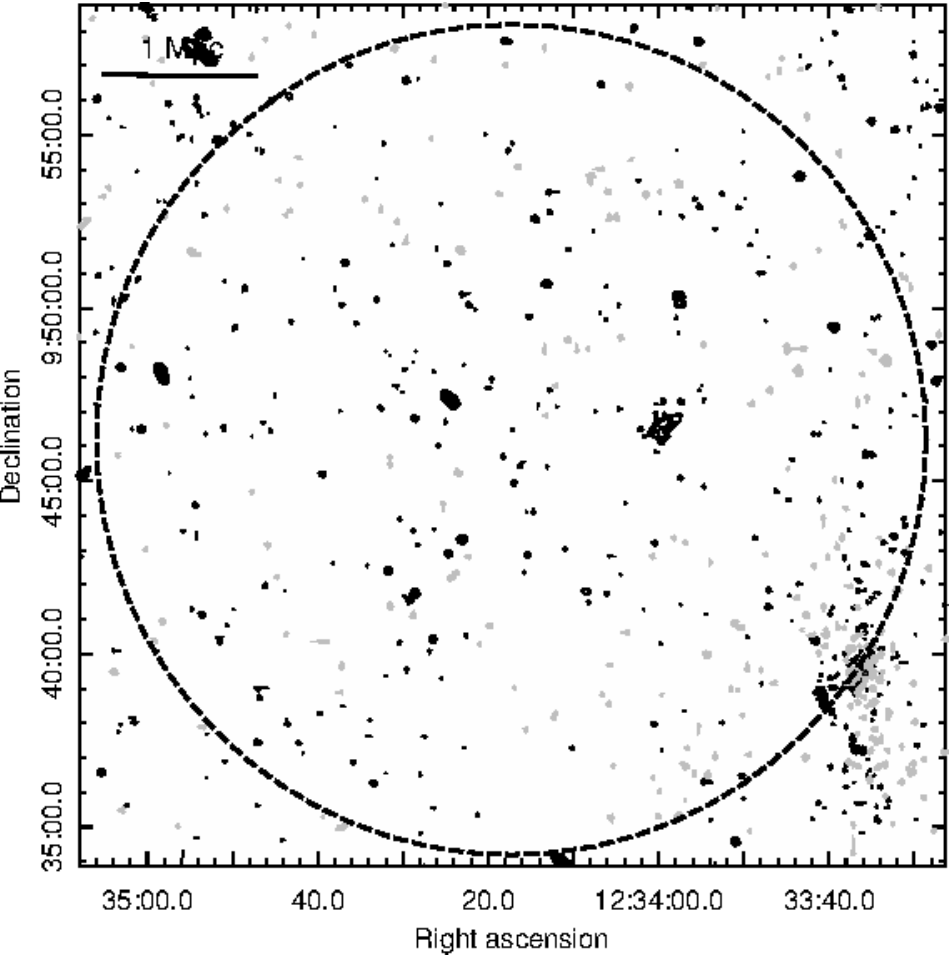}
\includegraphics[width=9cm]{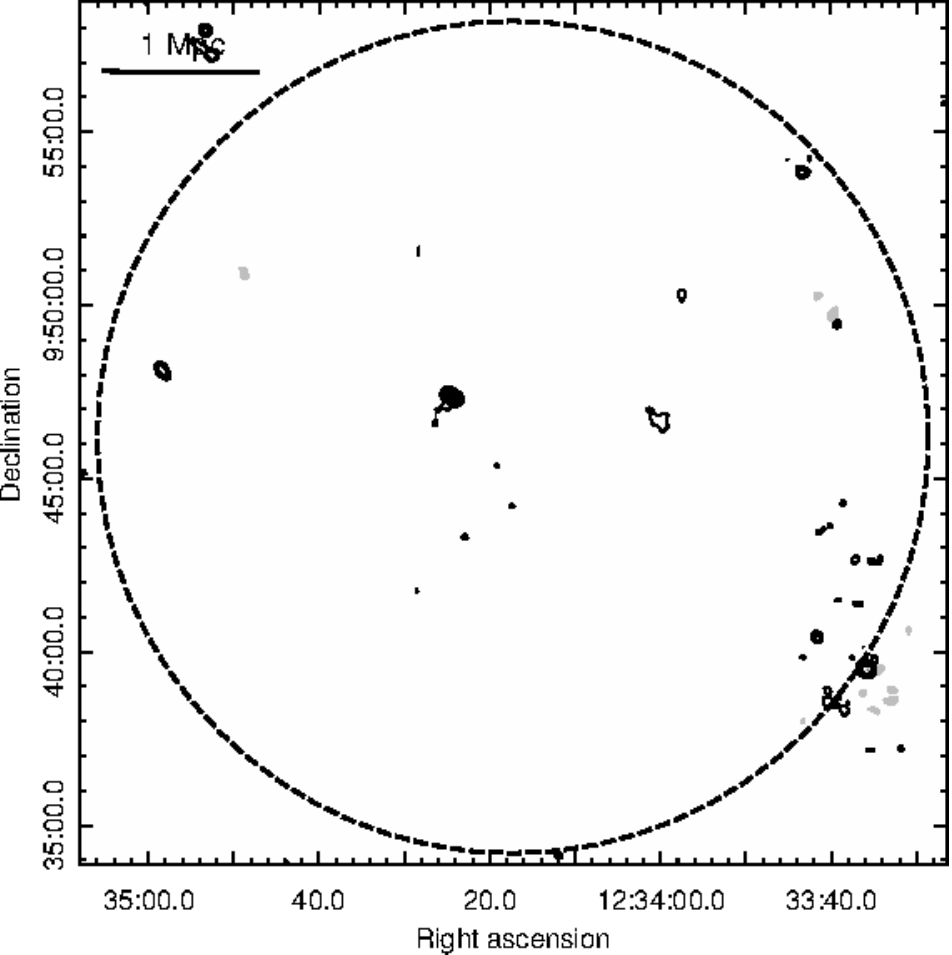}
\caption{{\bf Z5247 \,{\bf (RXC\,J1234.2+0947)}}: GMRT 610 MHz (left) and 235 MHz (right) images in 
contours. 
The contours are at $0.09\times(\pm1, 2, 4,...)$ mJy beam$^{-1}$ at 610 MHz 
and $2.4\times(\pm1, 2, 4,...)$ mJy beam$^{-1}$ at 235 MHz.  The
HPBWs at 610 and 235 MHz are $6.2''\times5.2''$, p. a. $56^{\circ}$
 and  $13.0''\times12.3''$, p. a. $71.4^{\circ}$, respectively. The 
$1\sigma$ levels in the 610 and 235 MHz images are 0.03 and 
0.6 mJy beam$^{-1}$, respectively. The circle has a radius equal to the virial 
radius for this cluster ($10.6'$).}
\label{appaz5247}
\end{figure*}

\begin{figure*}
\centering
\includegraphics[width=9cm]{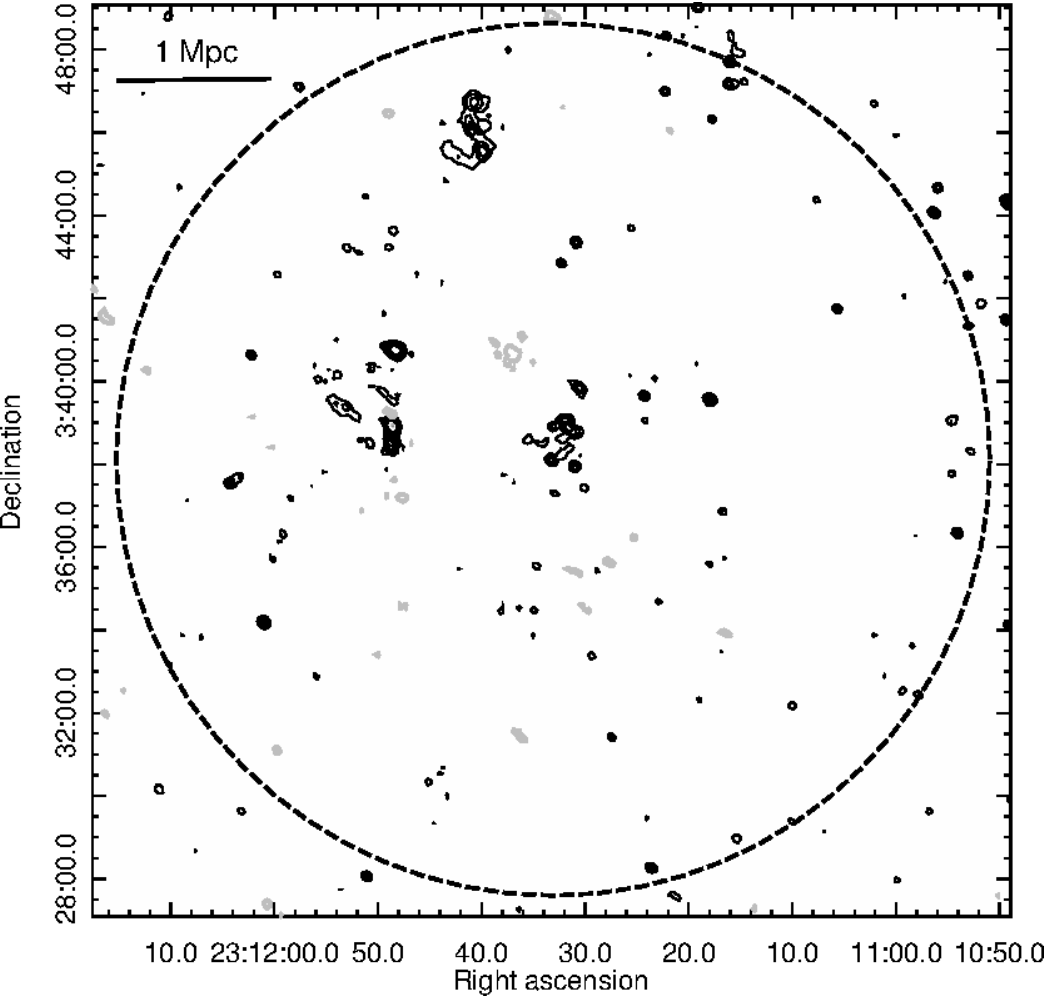}
\includegraphics[width=9cm]{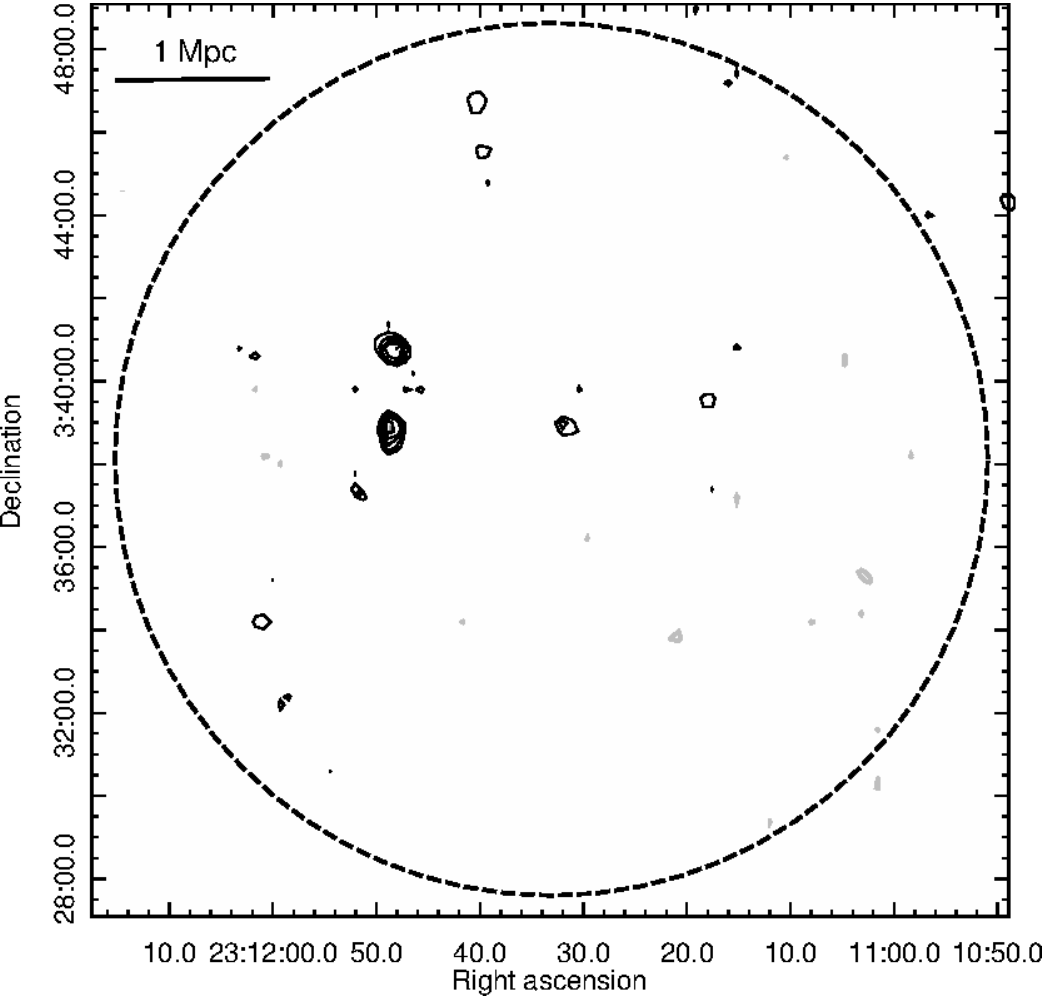}
\caption{{\bf A2552}: GMRT 610 MHz (left) and 235 MHz (right) images in 
contours. 
The contours are at $0.21\times(\pm1, 2, 4,...)$ mJy beam$^{-1}$ at 610 MHz 
and $2.7\times(\pm1, 2, 4,...)$ mJy beam$^{-1}$ at 235 MHz.  The
HPBWs at 610 and 235 MHz are $9.2''\times7.2''$, p. a. $50.4^{\circ}$
 and  $16.2''\times13.4''$, p. a. $9.6^{\circ}$, respectively. The 
$1\sigma$ levels in the 610 and 235 MHz images are 0.07 and 
0.7 mJy beam$^{-1}$, respectively. The circle has a radius equal to the virial 
radius for this cluster ($10.5'$).}
\label{apparxcj2311}
\end{figure*}

\begin{figure*}
\centering
\includegraphics[width=9cm]{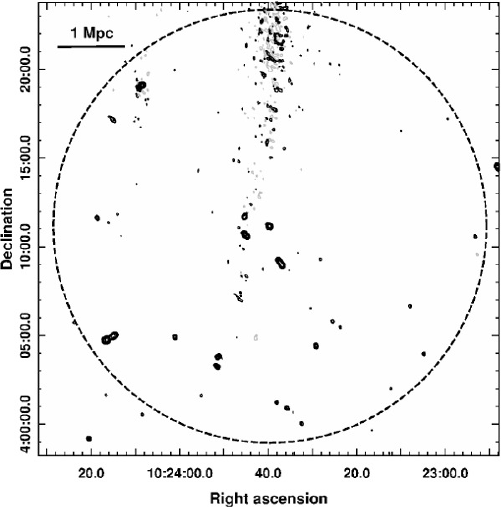}
\includegraphics[width=9cm]{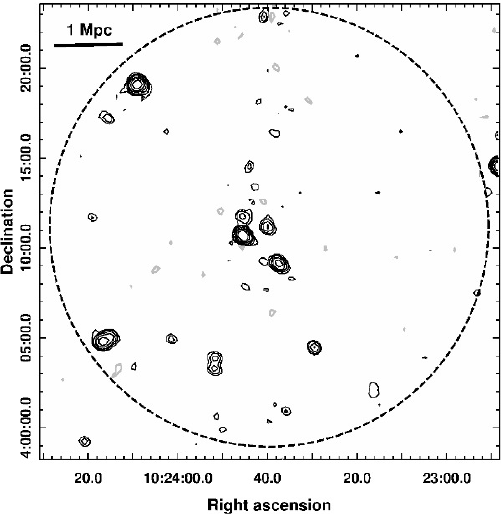}
\caption{{\bf Z3146 \,{\bf (ZwCL1021.0+0426)}}: GMRT 610 MHz (left) and 235 MHz (right) images in 
contours. 
The contours are at $0.6\times(\pm1, 2, 4,...)$ mJy beam$^{-1}$ at 610 MHz 
and $3.3\times(\pm1, 2, 4,...)$ mJy beam$^{-1}$ at 235 MHz.  The
HPBWs at 610 and 235 MHz are $8.9''\times7.1''$, p. a. $11.3^{\circ}$ and 
$26.7''\times21.1''$, p. a. $47.8^{\circ}$, respectively. The 
$1\sigma$ levels in the 610 and 235 MHz images are 0.09 and 
1.1 mJy beam$^{-1}$, respectively. The circle has a radius equal to the virial 
radius for this cluster ($12.2'$).}
\label{appaz3146}
\end{figure*}

\begin{figure*}
\centering
\includegraphics[width=9cm]{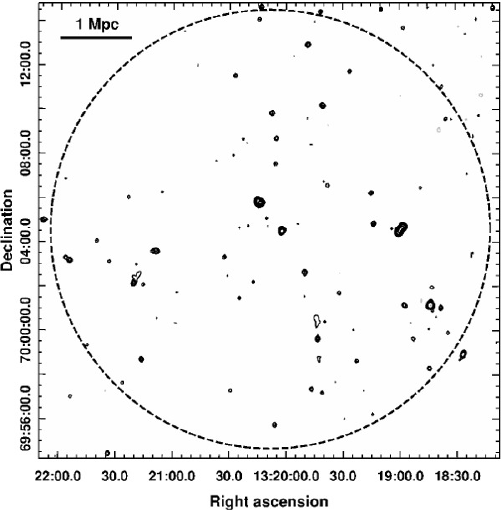}
\includegraphics[width=9cm]{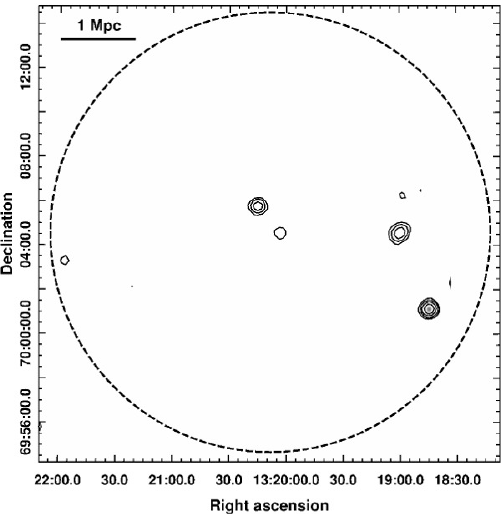}
\caption{{\bf A1722}: GMRT 610 MHz (left) and 235 MHz (right) images in 
contours. The contours are at $0.18\times(\pm1, 2, 4,...)$ mJy beam$^{-1}$ at 610 MHz 
and $5.0\times(\pm1, 2, 4,...)$ mJy beam$^{-1}$ at 235 MHz.  The
HPBWs at 610 and 235 MHz are $8.4''\times7.3''$, p. a. $-9.9^{\circ}$ and 
$26.2''\times24.3''$, p. a. $-48.6^{\circ}$, respectively. The 
$1\sigma$ levels in the 610 and 235 MHz images are 0.04 and 
0.9 mJy beam$^{-1}$, respectively. The circle has a radius equal to the virial 
radius for this cluster ($9.9'$).}
\label{appa1722}
\end{figure*}

\begin{figure*}
\centering
\includegraphics[width=9cm]{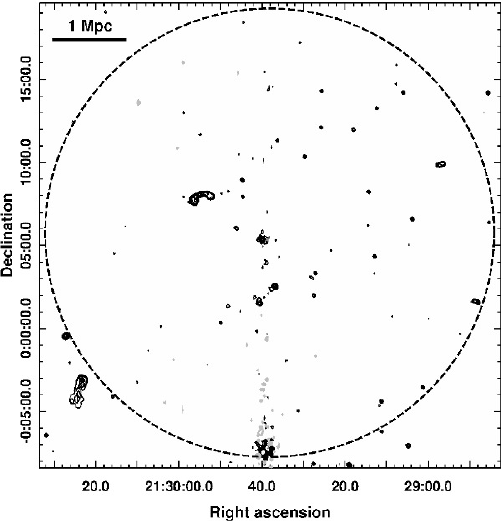}
\includegraphics[width=9cm]{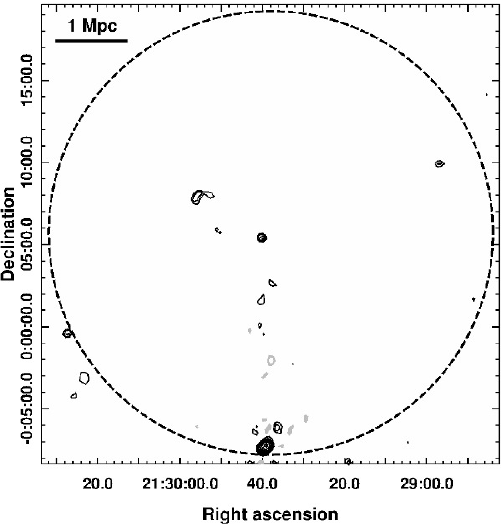}
\caption{{\bf RX\,J2129.6+0005}: GMRT 610 MHz (left) and 235 MHz (right) images in 
contours. The contours are at $0.3\times(\pm1, 2, 4,...)$ mJy beam$^{-1}$ at 610 MHz 
and $6.0\times(\pm1, 2, 4,...)$ mJy beam$^{-1}$ at 235 MHz.  The
HPBWs at 610 and 235 MHz are $7.9''\times6.3''$, p. a. $29.3^{\circ}$ and 
$14.5''\times11.3''$, p. a. $53.8^{\circ}$, respectively. The 
$1\sigma$ levels in the 610 and 235 MHz images are 0.08 and 
0.9 mJy beam$^{-1}$, respectively. The circle has a radius equal to the virial 
radius for this cluster ($13.5'$).}
\label{apparxj2129}
\end{figure*}

\begin{figure*}
\centering
\includegraphics[width=9cm]{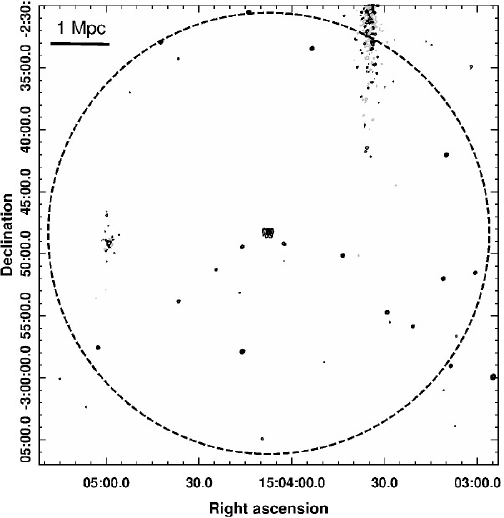}
\includegraphics[width=9cm]{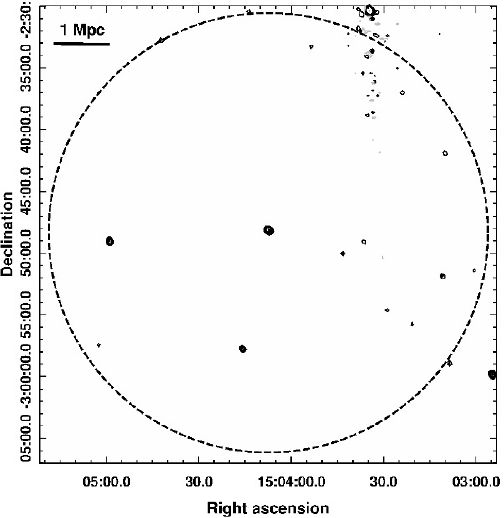}
\caption{{\bf RXC\,J1504.1-0248}: GMRT 610 MHz (left) and 235 MHz (right) images in 
contours. The contours are at $1.0\times(\pm1, 2, 4,...)$ mJy beam$^{-1}$ at 610 MHz 
and $6.0\times(\pm1, 2, 4,...)$ mJy beam$^{-1}$ at 235 MHz.  The
HPBWs at 610 and 235 MHz are $7.9''\times6.3''$, p. a. $29.3^{\circ}$ and 
$14.5''\times11.3''$, p. a. $53.8^{\circ}$, respectively. The 
$1\sigma$ levels in the 610 and 235 MHz images are 0.08 and 
0.9 mJy beam$^{-1}$, respectively. The circle has a radius equal to the virial 
radius for this cluster ($17.8'$).}
\label{apparxcj1504}
\end{figure*}

\begin{figure*}
\centering
\includegraphics[width=9cm]{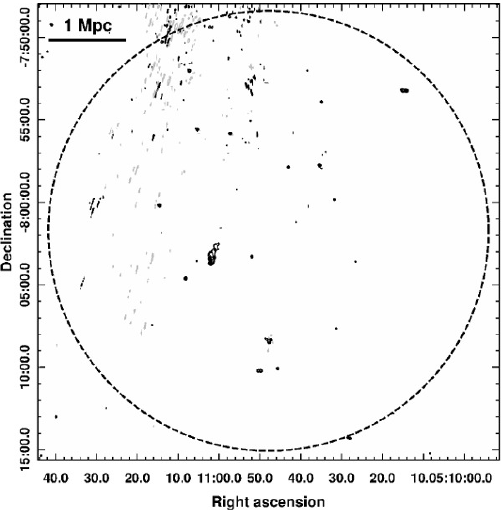}
\includegraphics[width=9cm]{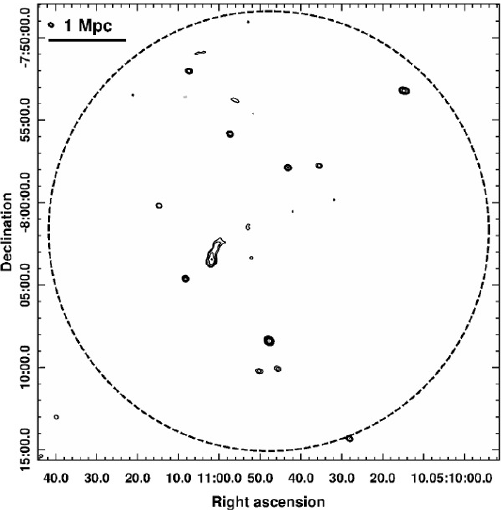}
\caption{{\bf RXC\,J0510.7-0801}: GMRT 610 MHz (left) and 235 MHz (right) images in 
contours. The contours are at $0.8\times(\pm1, 2, 4,...)$ mJy beam$^{-1}$ at 610 MHz 
and $6.0\times(\pm1, 2, 4,...)$ mJy beam$^{-1}$ at 235 MHz.  The
HPBWs at 610 and 235 MHz are $5.4''\times4.8''$, p. a. $41.1^{\circ}$ and 
$15.7''\times''13.1$, p. a. $69.6^{\circ}$, respectively. The 
$1\sigma$ levels in the 610 and 235 MHz images are 0.2 and 
1.2 mJy beam$^{-1}$, respectively. The circle has a radius equal to the virial 
radius for this cluster ($13.3'$).}
\label{apparxcj0510}
\end{figure*}

\end{appendix}

\end{document}